\documentclass[leqno,openbib]{article}
\usepackage{indentfirst}
\usepackage{amsfonts}
\usepackage{ntheorem}
\usepackage{amsmath}
\usepackage[doublespacing]{setspace}
\usepackage{sectsty}
\usepackage[norule,bottom]{footmisc}
\usepackage[justification=centering,textfont={sc},labelfont={rm}]{caption}
\usepackage{varioref}
\usepackage{natbib}
\usepackage{graphicx}
\usepackage{hyperref}
\usepackage{makecell}
\usepackage{amsmath}
\usepackage{geometry}
\usepackage{changepage}
\usepackage{float}
\usepackage[ruled,vlined]{algorithm2e}
\usepackage{amssymb}

\theoremindent\parindent
\makeatletter    
\renewtheoremstyle{plain}{\item[\hskip\labelsep\hskip-\parindent \theorem@headerfont ##1\ ##2\theorem@separator]}{\item[\hskip\labelsep\hskip-\parindent \theorem@headerfont ##1\ ##2\ (##3)\theorem@separator]}
\makeatother
\theoremheaderfont{\scshape}
\theorembodyfont{\rmfamily}
\theoremseparator{.}

\sectionfont{\normalfont\scshape\centering}

\subsectionfont{\itshape}
\makeatletter
\def\@biblabel#1{\hspace*{-\labelsep}}

\makeatother
\makeatletter
\renewcommand\@makefnmark{\mbox{\textsuperscript{\normalfont\@thefnmark}}}
\renewcommand\@makefntext[1]{\indent\makebox[2.5em][r]{\@thefnmark.\,}#1}
\if@titlepage
  \renewenvironment{abstract}{      \titlepage
      \null\vfil
      \@beginparpenalty\@lowpenalty
      \begin{center}        \@endparpenalty\@M
      \end{center}}     {\par\vfil\null\endtitlepage}
\else
  \renewenvironment{abstract}{      \if@twocolumn
      \else
        \small
        \begin{center}        \end{center}      \fi}
      {\if@twocolumn\else\endquotation\fi}
\fi
\renewcommand\thetable{\@Roman\c@table}

\renewcommand\thefigure{\@Roman\c@figure}

\makeatother
\labelformat{equation}{(#1)}
\begin{document}

\title{Fast Simulation-Based Bayesian Estimation of Heterogeneous and Representative Agent Models using Normalizing Flow Neural Networks}
\author{%
\textsc{Cameron Fen\thanks{Cameron Fen is PhD student at the University of Michigan, Ann Arbor, MI, 48104 (E-mail: camfen@umich.edu. Website: cameronfen.github.io.). The author thanks David Childers, Eric Jang, and Florian Gunsilius for helpful feedback and Alisdair McKay, Toni Whited, Emil Lakkis, Mohammed Ait Lahcen, and Jeppe Druedahl for code. All errors are my own.}}}
\maketitle

\vspace*{-2cm}
\begin{abstract}
This paper proposes a simulation-based deep learning Bayesian procedure for the estimation of macroeconomic models. This approach is able to derive posteriors even when the likelihood function is not tractable. Because the likelihood is not needed for Bayesian estimation, filtering is also not needed. This allows Bayesian estimation of HANK models with upwards of 800 latent states as well as estimation of representative agent models that are solved with methods that don't yield a likelihood--for example, projection and value function iteration approaches. I demonstrate the validity of the approach by estimating a 10 parameter HANK model solved via the Reiter method that generates 812 covariates per time step, where 810 are latent variables, showing this can handle a large latent space without model reduction. I also estimate the algorithm with an 11-parameter model solved via value function iteration, which cannot be estimated with Metropolis-Hastings or even conventional maximum likelihood estimators. In addition, I show the posteriors estimated on Smets-Wouters 2007 are higher quality and faster using simulation-based inference compared to Metropolis-Hastings. This approach helps address the computational expense of Metropolis-Hastings and allows solution methods which don't yield a tractable likelihood to be estimated. 
\begin{singlespace}

JEL Codes: E27, E37, E47, C45, C11, C15\newline
Keywords: Neural Networks, Bayesian Inference, Dynamic Estimation, Simulation-Based Estimators
\end{singlespace}
\end{abstract}


\thispagestyle{empty}\setcounter{page}{0}%
\newpage 

\section{Introduction}
There are significant barriers to Bayesian or maximum likelihood estimation (MLE) in many situations outside one standard approach of estimating after perturbation. In spite of this, Bayesian methods have advantages. They allow the use of informative priors, perhaps derived by microdata to guide estimation and provide distributional information on parameters. They give distributional insight on parameters. In the case of MLE, it efficiency advantages over alternatives like method of simulated moments and modeling advantages over maximum simulated likelihood. 

Discussing the barriers, first, MH-MCMC estimation of Bayesian posteriors, particularly of slow state-of-the-art models, is computationally inefficient. For example, even in a small HANK model, the size of the latent state space could be in the thousands making a filtering likelihood approach to Bayesian estimation intractable. For this reason, most heterogenous agent models are calibrated \citep{liu2021full}, but if they are estimated in a likelihood manner, state space dimension reduction techniques are used at the expense of accuracy \citep{ahn2018inequality}. Second, many popular solution methods, like value function iteration or projection, don't yield complementary likelihood functions so conventional methods using Metropolis-Hastings Markov Chain Monte Carlo (MH-MCMC) and MLE are off the table. 

Discussing the first barrier: the problem with heterogeneous agent model estimation is that the Kalman filter is a computational bottleneck when the state space is too large. Since heterogeneous agent models often divide up one or more distribution state variable into many quantiles, the size of the latent state space could be in the 1000s. Since Kalman filtering's computational complexity scales with matrix multiplication, filtering something that is approximately 20x larger than the 40 parameter latent space in \citet{smets2007shocks} model takes approximately 4500x more compute. This means the traditional MH-MCMC or even maximum likelihood combined with a filtering approach may become computationally intractable, even for small heterogeneous agent models. The approach I propose estimates the posterior from simulations only, allowing one to avoid filtering and the likelihood function in general. Thus, this approach can estimate models with almost unbounded latent state space size with small increases in computational cost.    

Likewise, discussing the second barrier: conventional approaches have downsides if the solution method doesn't yield a complementary likelihood function. If one cannot estimate via perturbation, Bayesian and MLE estimation cannot be done with conventional MH-MCMC or even MLE. Thus options with less attractive qualities are used. In this case, solved models are either calibrated, estimated via method of simulated moments (MSM), or estimated in a simulated maximum likelihood fashion. Each approach has its own drawbacks. Calibration is not an algorithmic approach so it lacks both rigour and quantification tools like standard errors. MSM is not efficient unless the global identification criterion is met, which is both unverifiable and unlikely. Adding in measurement error for maximum simulated likelihood makes it less realistic as practitioners often want to assume the error comes from economic shocks and not measurement. Additionally, none of these alternatives can handle incorporating posterior distributions or prior information. 

I propose a method from machine learning that allows Bayesian and full information estimation of models without a likelihood function. This also addresses filtering computational bottlenecks since avoiding the likelihood function avoids the use of filtering. Since the approach is likelihood-free, one can Bayesian/MLE estimate models solved via projection and value function iteration. This includes non-analytic models with kinks that often require these solution approaches. This approach allows the estimation of heterogeneous agent models whose large latent state spaces make filtering intractable without dimension reduction, even when the likelihood is available. Although the method is more general, this paper focuses on the case of dynamic macroeconomic structural models in a Bayesian setting. Finding the mode of the posterior with a uniform prior allows one to derive MLE estimation from the Bayesian procedure as well. Econometrically, this approach extends \citet{kaji2020adversarial} to the Bayesian setting and when there is a time series component that is not iid. 

I next will discuss briefly the Sequential Neural Posterior Estimation (SNPE) algorithm \citep{greenberg2019automatic}, \citep{tejero-cantero2020sbi} that can perform this likelihood-free Bayesian and MLE estimation. SNPE uses a model little used in economics, the normalizing flow \citep{rezende2015variational}, which is a powerful conditional density estimator. Flows facilitate Bayesian estimation by learning the posterior conditional distribution $P(\theta|x)$ trained on samples from the joint, $x,\theta \sim P(x,\theta) = P(x|\theta)P(\theta)$. One can think of SNPE as an extension of the \citet{kristensen2012estimation} approach to likelihood estimation, where they simulate data from the joint via the prior and the likelihood simulations. Then they use a kernel density estimator (KDE) to estimate the likelihood function. In my case, I use a normalizing flow to replace their KDE, which allows me to extend their result to full Bayesian inference and accommodate data with a latent state space structure. The benefits of SNPE mainly stem from the use of the flow over traditional density estimators like the KDE. There are also alternative methods like SNRE \citep{durkan2020contrastive} which is discussed in the appendix and simulation-based inference combined with variational inference \citep{glockler2021variational}.  

The general approach around simulation-based inference has gained popularity in many fields. The approach is often used for Bayesian inference in machine learning \citep{durkan2020contrastive}, \citep{greenberg2019automatic}. The approach has been used in neuroscience \citep{boelts2021flexible} and ecology \citep{dinapoli2021approximate}. The technique has also become one of the leading estimation methods in many fields of physics \citep{brehmer2021simulation}, \citep{cranmer2020frontier}. 

\section{Literature Review}
\label{lit_review_section}
This literature review will cover three topics: simulation-based methods in economics, the literature on solving dynamic models, particularly with kinks and discontinuities, and solution and estimation of heterogenous agent models. I will discuss MH-MCMC and machine learning background in the following sections.   

\subsection{Simulation-Based Models in Economics}
I will give an overview of current simulation-based estimation and related approaches. Most of the simulation-based likelihood and inference approaches are inspired by an approach like method of simulated moments (MSM) \citep{mcfadden1989method}, \citep{pakes1989simulation}, \citep{duffie1990simulated}. Due to the difficulty of verifying the global identification criterion there is a large interest in developing more efficient and robust estimators where this is not a problem. Techniques like maximum simulated likelihood, attempt to address this \citep{lerman1981use}, although these techniques both have simulation bias \citep{hajivassiliou1997some}, \citep{haan2006estimation} and typically require the use of measurement error. There has also been work on using techniques from simulation-based likelihood estimation to solve dynamic models, particularly in industrial organization \citep{keane1994solution}. There is also literature on efficient simulation techniques using both indirect inference \citep{smith1993estimating}, \citep{gourieroux1993indirect}, and efficient method of moments \citep{gallant1996moments}. 

Simulation-based techniques in economics are efficient when the model is well specified, so much of recent work has been done to improve robustness in misspecification. One such avenue is efficient method of moment estimators using a spectrum of moments so the data is guaranteed to be in the support of distributions spanned by the moments\footnote{There is also work in machine learning attempting to perform robust and efficient moment estimation by using a spectrum of moments, most well known is the Generative Moment Matching Network \citep{li2015generative}} \citep{carrasco2007efficient}, \citep{altissimo2009simulated}. However, these methods have a difficult time dealing with models with latent time structure. Another approach is approximate Bayesian inference \citep{rubin1984bayesianly} to estimate models solved via value function iteration techniques, to estimate these models. However, in a well specified model, this approach is only approximately Bayesian, unlike the set of algorithms I propose. 

One could even consider a particle filter a simulation-based technique for deriving a likelihood function given an intractable integral \citep{fernandez2007estimating}, but this approach only works with dynamic models that yield complementary state space representations and requires a likelihood function at each point in time.  Although the literature is more sparse, there are papers that also perform simulated Bayesian inference \citep{flury2011bayesian} \citep{herbst2014sequential}, extending particle filtering for Bayesian inference. Other machine learning approaches use techniques that have both robustness and efficiency guarantees like GANs. Most relevant to this paper, \citet{kaji2020adversarial} whose GAN approach to structural modelling open the door to robust and near efficient point estimation using only simulations from the model and not a likelihood. 

\subsection{Solving Dynamic Models with Kinks}
I will next discuss approaches to solve dynamic models, particularly models with kinks and nonlinearities.  

There are a variety of ways to solve dynamic macroeconomic models: perturbation, projection, and value function iteration \citep{fernandez2016solution}, \citep{judd2017solve}, of which only perturbation yields likelihood functions that allow Bayesian and MLE estimation using conventional techniques. Perturbation works well when the policy function looks like a linear or low degree polynomial function. For methods that are highly nonlinear, with large shocks \citep{terry2017alternative}, or even non-analytic, projection and value function iteration should be used. For a large set of models, almost anything that is solved via value function iteration \citep{hennessy2007costly}, heterogeneous agent models with intractable likelihoods \citep{kukackaanon}, and most projection models the likelihood is intractable forcing the use of less efficient point estimate methods that have weaknesses compared to Bayesian and MLE estimation.  

In order to estimate non-perturbation models, one has to resort to simulation-based inference approaches. As such, most of the time, these models are not estimated in a Bayesian or even full information manner and practitioners resort to, for example, method of simulated moments. 

As an illustration of problems faced by a perturbation approach, below is an example of a s-S model policy function in \citet{caballero2007price}: 

\begin{figure}[htbp]
    \centering
    \includegraphics[width=10cm]{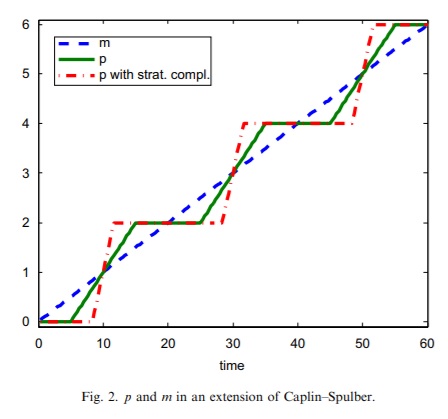}
    \caption{s-S model}
    \label{fig:s-S model}
\end{figure}

s-S models posits a fixed cost to changing a state variable which typically leads to kinks in the policy function. Often times modeling inventory or capital requires this assumption. The above example uses s-S costs with respect to menu prices assuming a linearly increasing money supply (blue). This is a modification of \citet{caplin1987menu}, which assumes there is a spectrum of firm prices from the s-S trigger point to zero. \citet{caballero2007price} assume there is a spectrum of firms prices that only covers a portion of the range from the s-S trigger to zero. As the money supply increases linearly, at some time periods there will be no firms adjusting due to the fixed costs and at some time periods an above average number of firms will adjust to keep up with the price level. This is shown by the behavior in the the green and red line show. The kinks are where the policy function is non-differentiable and shows the transition where no firm is adjusting to when some firms are adjusting prices or vice versa. Thus a perturbation approach, which estimates a Taylor series approximation of the policy function, will not work as the policy function is non-analytic. This implies that one must solve the model via something like value function iteration, which leads to estimation problems mentioned above. 

Papers which estimate s-S models include \citet{arrow1951optimal} and \citet{caplin1987menu}. In particular, \citet{khan2008idiosyncratic} assumes an s-S model for heterogeneous agent model for investment in capital. \citet{house2004ss} model the purchasing behavior for used durable goods given a fixed cost in addition to the price of the good. In particular, the good is priced under adverse selection \citep{akerlof1978market}. Likewise in finance there is a large literature estimating structural models \citep{he2021relative}, \citep{taylor2013ceo}. Like the s-S models, this literature often includes fixed adjustment costs or other approaches that yield kinks in the policy function. \citet{terry2022information} finds a negative externality that comes regulation of public company disclosure. Although they find current disclosure regulations are low, if one is forced to disclose accurately, one will mis-invest so that the investments generate results that more closely resemble the the desired upon false disclosure results.  

\subsection{Heterogeneous Agent Models}
I will now discuss the literature around full information estimation of heterogenous agent models. Much of the methodological interest in heterogeneous agent models is improving the speed of solution methods \citep{auclert2021using}, \citep{winberry2018method}, \citep{khan2008idiosyncratic}. However the computational bottleneck for estimation problems is not solution speed, but the intractability of the Kalman filter on large latent spaces. For this reason, while there is a large literature on heterogeneous agent model \citep{auclert2021demographics}, \citep{ottonello2020financial}, \citep{caglio2021risk}, most practitioners resort to calibration because of this and other barriers to estimation. There have been approaches like \citet{ahn2018inequality}, who propose a dimension reduction technique to reduce latent state space size to be manageable for filtering. However this approaches throws away information while still adding overhead compared to a pure simulation-based approach. \citet{liu2021full} and \citet{parra2020estimation} perform maximum likelihood estimation using micro data along with macroeconomic aggregates. \citet{parra2020estimation} estimate an Aiyagari model \citep{aiyagari1994uninsured}, \citep{huggett1993risk} and propose a diagnostic that finds that a sizable number of parameters are not well identified and they suggest calibrating those parameters. For my results, the choice to use uniform priors for all the estimation problems faces the same documented identification problem, however, uniform priors are still used for the sake of estimation transparency. 

While there is a large literature dealing with simulation based estimators, both macroeconomists that work with heterogeneous agent and representative agent models have expressed a need to reduce the barriers to MLE and Bayesian estimation. The SNPE and related algorithms can help mitigate these barriers.  

\section{Background on Simulation Neural Posterior Estimation (SNPE) and Metropolis-Hastings}
This section will discuss the background of the SNPE estimator and MH-MCMC. First, I will intuitively discuss the basics behind Bayesian estimation. Then I will move to a brief discussion of normalizing flows, which is a density estimator used to estimate the posteriors from samples drawn from the joint distribution. I will discuss the SNPE algorithm and finally conclude with some caveats and pathological examples.  

When discussing the approach of simulation-based inference, it is useful to lay down a few definitions. The true data comes from the underlying data generating process. For macroeconomics, it would be economic data like output, consumption, etc. The simulator is the model being estimated, for this paper: a dynamic macro model. The only demand simulation-based inference puts on simulators is to be able to simulate data that has a 1-1 correspondence to the covariates in the true data. 

\subsection{Bayesian Basics}
\label{BB}
In this subsection, I will discuss the general paradigm of Bayesian estimation, then I will discuss MH-MCMC. 

Bayesian estimation attempts to find the posterior given the likelihood and prior of a model using Bayes' rule: 
$$
    P(\theta|x) = \dfrac{P(x|\theta)P(\theta)}{P(x)}
$$
Here $P(\theta|x)$ is the posterior, $P(x|\theta)$ is the likelihood, and $P(\theta)$ is the prior. All approaches that perform Bayesian inference, ranging from MH-MCMC, variational inference, to simulation-based inference concern techniques for calculating $P(\theta|x)$ without calculating the partition function, $P(x)$.  

The predominant technique for performing Bayesian inference in economics is to perform MH-MCMC \citep{herbst2015bayesian}, which I will describe next. In MH-MCMC, one concerns oneself with likelihood ratios between parameter values. For example, if one knows both the prior and the likelihood for any given point, one knows the relative likelihood of being in any point versus any other. One point in parameter space may be twice as likely to be visited in the posterior as another point, even if the actual probabilities are not known. Since one knows the relative probabilities, one can design a random walk so that a walker visits probabilities equivalent to the ratios of there probabilities. One way to do that is if a walker can choose to compare any point in the posterior to the point the walker is currently at, the walker will move to the second point with the likelihood according to the ratio of the unnormalized probabilities. If the second point is half as likely to be in the posterior, the walker will move with probability one half and probability one half stay in place.  If the second point is more likely to be in the posterior, say twice as likely, the walker will move to the second point for sure, with the understanding that if the walker randomly selected to move back, the move back would be now with half the probability. 

In this way, the walker moves across the space of the posterior with relative frequency according to $P(x|\theta)P(\theta)$ ratio between points, which ultimately means that one is sampling from $P(\theta|x)$ without calculating the normalizing partition function. In practice, MCMC is a little more complicated, as often the parameter space is unbounded, and one can't sample the entire space with equal probability if the space is unbounded. Thus this involves the use of a proposal distribution and the use of importance sampling (see appendix for more information). 

Alternatives like variational inference have been proposed which speed up the inference at the cost of some bias \citep{wainwright2008graphical}, but it is beyond the scope of this paper. The approach I propose, simulation-based inference, does not require knowing the likelihood, $P(x|\theta)$, which often cannot be derived with commonly used solution methods, and scales computationally much better than MH-MCMC for larger state-of-the-art models, like HANK models. The backbone of the SNPE approach I proposes samples points from the joint distribution $P(x,\theta)$, then one estimates the posterior $P(\theta|x)$ using a conditional density estimator. SNPE uses a normalizing flow as a density estimator which has a series of advantages over traditional KDEs. The normalizing flow and it's advantages will be the next topic.    

\subsection{Normalizing Flows}
section{Normalizing Flows} \label{flows}
In this section, I will discuss how a particular normalizing flow, the neural autoregressive flow \citep{huang2018neural}, is structured. Then I will discuss how to use the change of variables formula for a random variable to derive a likelihood of a sample under the flow. I will also attempt to highlight advantages flows have over KDEs that make the SNPE algorithm more powerful that the tradional \citet{kristensen2012estimation} approach.   

A flow is a composition of individual bijectors. Thus the first task is to define these intermediate changes in measure, $y^a,y^b...y^z$, so that they can be composed with each other to form a invertable function. For example the output of $f_a$ becomes the input to $f_b$ giving the total composition much more flexibility. This is not trivial. For example take a polynomial regression. If one knows the input to this regression, one can calculate the output. However if one knows the output, the roots of a polynomial are not generally solvable and so polynomials aren't invertable transformations. However one set of invertable transformations are affine operations. Given $y^i,y^j \in R^N$ and $n$ going from 1...N indexing each element in the vector $y^i$:

\begin{equation}
    \forall n, y^j_n = \beta^0_n + \beta^1_n*y^i_n
\end{equation}

This is trivially invertable as one can derive $y^i$ from $y^j$ by shifting the negative of $\beta^0$ and scaling by the inverse of $\beta^1$. While one can still add nonlinearities like a logit link function between bijective transformations, this still lacks expressivity. The standard way to extend this model is to allow for the shifts and scales to be dependent on at least some of the inputs. For example, earlier elements $y^i_{a:l}$ are only modified in an directly invertable manner (ie $y^i_{a:l} = \beta^0 + \beta^1*y^j_{a:l}$), then condition the shifts and scales for the rest of the elements $y^i_{l:N}$ on inputs, $y^i_{a:l}$, that weren't transformed: 
\begin{equation}
    \forall \iota \in 2:N, y^j_\iota = \beta^0_\iota(y^i_{a:\iota-1}) + \beta^1_\iota(y^i_{a:\iota - 1})*y^i_\iota
\end{equation}
$\beta^0(.)$ and $\beta^1(.)$ are typically differentiable and flexible functions--typically feed-forward neural networks conditioned on earlier elements of the vector, that now output dynamic "psuedo" parameters $\beta^0$, $\beta^1$ that depend on the conditioning variables, $y^i_{a:\iota-1}$. Next I will illustrate how a neural autoregressive bijector is defined that I use in my paper. Make the first element, $y^j_1$ the identity or affine map, then the second element $y^j_2$ an affine transformation in $y^i_2$ with psuedo-parameters $\beta_0(y^i_1)$ and $\beta_1(y^i_1)$ dependent only on $y^i_1$. The third element, $y^j_3$, has psuedo-parameters then dependent only on $y^i_{1:2}$ and $y^i_3$ enters only in an affine manner. 

$$y^j_\iota = \sigma(\beta^0_\iota(y^i_{a:\iota-1}) + \beta^1_\iota(y^i_{a:\iota - 1})*y^i_\iota)$$

To make this model nonlinear, one typically adds a link function $\sigma$ which is usually a Leaky ReLU \citep{xu2015empirical}. A Leaky ReLU is two lines of different slope intersecting at zero.  This kink is often enough to make neural networks universal approximators of continuous functions \citep{cybenko1989approximation}.  

Since $\sigma$ is invertable as well as the rest of the flow, one can recover the inputs $y^i$ knowing only all elements in $y^j$\footnote{The procedure to recover $y^i_2$ from all the $y_j$'s is to know $y^i_1$ from the affine first equation. Then recover the affine parameters for the $y^i_2$ equation (which is only conditioned on $y^i_1$). Then one can invert $y^j_2$ to get $y^i_2$ knowing the affine parameters. Now that one knows $y^i_{1:2}$ one can repeat the procedure to get $y^i_3$, etc.}. This makes the bijector invertable in practice. This invertability is another advantage normalizing flows have over KDE. One can sample from them and calculate the pdf of any sample without the use of rejection sampling or other computationally heavy techniques. This will be exploited in the SNPE algorithm as the flow will be used both as a density estimator and as a proposal distribution for generating $\theta$ samples.

Additionally since each affine transformation is a function of variables that have an index smaller than the index in question, the Jacobian is lower triangular and the determinant is just the product on the diagonal. This makes the change in variable formula scale linearly with the number of parameters.  

Furthermore, if one stacks multiple bijections on top of one another, one can permute the order of input elements ($y^i_{1:N}$) with a bijective operation so that different bijectors have different conditioning relationships among variables allowing the flow to universally approximate any distribution. If the functions for $\beta^0(.)$ and $\beta^1(.)$ are universal approximators (ie neural networks), one can show that a stack of bijectors (along with permutations), is also a universal approximator and thus can approximate any change of variable arbitrarily well. This will be discussed and proved in the theoretical section in the appendix following \citep{huang2018neural}. 

One can also condition $\beta^0(.)$ and $\beta^1(.)$ with additional arbitrary conditioning variables to form a flexible conditional density estimator. This ability to condition is one advantage it has over a KDE. Since the problem is to estimate the density of $\theta$ conditioned on $x$, $\theta$ will take the role of $\{y^a_1..y^a_n\}$. $X$ will be the true data and analogue of simulated data which will be differentiated by a lower case $x$. Conditioning variable $x$ and real data $X$ will enter as additional conditioning varibles in the flow. In particular they will be included as conditioning variables in the neural networks that produces $\beta^0(.)$ and $\beta^1(.)$. Thus, for example $beta^1_5$ will be conditioned on $y^i_1...y^i_4$ as well as $x$/$X$, the simulated/real data.  

Since these models can model any conditional distribution, and can calculate the likelihood of any sample point in the target space, one can fit this model on samples from arbitrary continuous densities and have the guarantees that come with a well specified maximum likelihood problem. Thus, normalizing flows are density estimators that are robust to misspecification and asymptotically efficient due to estimation with maximum likelihood. 

Next I will discuss the change of variable formula for a flow. Given a density, $p(y^a)$ with random vector realizations $y^a$; a normalizing flow is a function, $f(y^a)$ mapping $y^a$ to a target random variable $y^z$, which has density, $q(y^z)$ such that the probability $q(y^z)$ is related to the probability $p(y^a)$ via:
\begin{equation}
    q(y^z) = p(y^a)|det \frac{df^{-1}}{dy^a})| = p(y^z)|det \frac{df(y^a)}{dy^a}|^{-1}
\end{equation}
This formula is simply the change of variable formula one learns in an introductory PhD econometrics class, with the first term $p(y^a)$ representing the measure in the base distribution, and the Jacobian representing the change in measure due to the transformation to the target distribution $q(y^z)$. In the appendix, I show in more detail how to construct a flow so that it is both invertable and can approximate any continuous distribution. Assuming both these things, one can use the change of variable formula to derive the likelihood that a sample point $y^z$ was generated by the flow \citep{rezende2015variational}. One uses the flow to transform the $y^z$ to $y^a$ and then one can use the change of variable formula to calculate $q(y^z)$. One can then perform maximum likelihood by modifying the parameters of the neural networks that govern the psuedo-parameters ($\beta^0_i(.)$ and $\beta^1_i(.)$ in each layer to maximize $q(y^z)$ for $y^z$ samples in the data. 

\subsection{Sequential Neural Posterior Estimation}
\label{SNLE and SNPE}
First, I will discuss the procedure underlying \citet{kristensen2012estimation}, which is the same approach as the SNPE algorithm. Then I will discuss the role of the normalizing flow in extending the \citet{kristensen2012estimation} approach as well as multi-round inference.  
I proceed by simulating $\theta$ from the prior. Then one simulates $x$, simulated data, from the model likelihood, $P(x|\theta)$. Concatenating the two simulations together gives samples from the joint distribution $x,\theta \sim P(x,\theta)$. With these samples one can use the normalizing flow density estimator to estimate $P(\theta|x)$. Setting the conditioning variable in the flow, $x$, to be the real data $X$ (note capitalized), allows one to have an estimate of the posterior.  

One typically proceeds with SNPE in a multi-round fashion where a proposal distribution which has more alignment with the posterior takes the place of the sampling from the prior. Since the true posterior and the true prior might have limited overlap, to reduce variance, it's often important to estimate the posterior in multiple rounds. In the first round, one samples from the prior. In later rounds, one samples from the current estimate of the posterior and performs importance sampling to correct from the distribution shift. More mathematically, given a proposed prior of $p'(\theta)$, true prior of $p(\theta)$, the posterior, when trained on this data, will have to be importance sample adjusted by the factor of $\dfrac{p(\theta)}{p'(\theta)}$, to account for sampling from a distribution that isn't the prior. \citet{greenberg2019automatic} perform the estimation in one step by recognizing the adjusted distribution $p'(\theta|x) = p(\theta|x)\dfrac{p'(\theta)}{p(\theta)}$, where $p'(\theta|x)$ is the posterior obtained by sampling from a proposal distribution different then the prior. Then if a normalizing flow $f_\phi(\theta|x)$ is estimated in place of $p(\theta|x)$ in the above equation, a fully flexible $f_\phi(\theta|x)$ will return a unbiased estimator of the posterior. Importance sampling is more thoroughly discussed in the appendix.  

\begin{algorithm}[H]
\SetAlgoLined
\textbf{Input:} Simulator $p(x|\theta)$, prior $p(\theta)$, data $x_0$, flow $f_\phi(x|\theta)$, Rounds R, Samples S\;
\textbf{Initialize:} Posterior $p^{(0)}= p(\theta)$, data set D = \{\}\;
\For{$i\gets 1$ \KwTo $R$}
{
    Sample $\theta^{(n)}\sim p^{(i-1)}$ for $n = 1...S$ with Monte Carlo\;
    Simulate $x^{(n)} \sim P(x|\theta^{(n)})$ for $n = 1...S$\;
    Concatenate data $D = D \cup \{x^{(n)},\theta^{(n)}\}^S_{n=1}$\;
    \While{$d_\phi(x,\theta)$ not converged}
    {
        Sample $\{x^{(i)},\theta^{(i)}\}^B_i \sim D$ from D\;
        Train $f_\phi(\theta|x)\dfrac{p^{(i-1)}}{p(\theta)}$ on $\{x^{(i)},\theta^{(i)}\}^B_i$\;
    }
    Update posterior $p^{(i)}\propto f_\phi(\theta|x)$\;
}
\caption{SNPE Algorithm}
\end{algorithm}

\subsection{Properties of SNPE}
This section will briefly relay the proof for why the SNPE algorithm converges to the Bayesian posterior in the infinite Monte Carlo sampling limit.  

Proposition 1 from \citet{papamakarios2016fast} proves that if $\theta$ is sampled from a proposal distribution $p'(\theta)$ with the true prior $p(\theta)$ and the likelihood is sampled from $p(x|\theta)$ than a normalizing flow $f_\phi(\theta|x)$ that maximizes the likelihood of the simulated data will be proportional to $\dfrac{p'(\theta)}{p(\theta)}p(\theta|x)$, ie:
\begin{align}\label{eq:7}
    f_\phi(\theta|x) \propto \dfrac{p'(\theta)}{p(\theta)}p(\theta|x)
\end{align}

provided that the true parameterization is contained in the set of parameters of the normalizing flow. This restriction implies that the posterior is continuous and at least $L^2$ for instance because the flow can only approximate continuous and $L^2$ distributions.  

The intuition behind the proof is that given enough samples from $x,\theta$, maximum likelihood of estimation will converge to the distribution that minimizes the KL divergence between the distribution the sample comes from and the normalizing flow parameterization. Thus if the flow has parameters that would set the KL-divergence to 0, this would be the actual distribution $\dfrac{p'(\theta)}{p(\theta)}p(\theta|x)$ that is the data generating process. When fitting distributions, since \ref{eq:7} implies the distribution that $f$ will converge too, if one estimates $f_\phi(\theta|x)\dfrac{p'(\theta|x)}{p(\theta)}$ on the data, $f$ will converge to the posterior. The importance weights are inverted since this function is trained on data rather than the data being reweighted by the weights. See importance sampling discussion in appendix for details. This proof goes hand in hand with the proof that a flow is a universal approximator (proof in appendix), as given enough samples and a large enough flow, the true posterior will be arbitrarily close in a KL divergence sense to the best parameterization of the flow. 

\subsection{Caveats and Pathological Examples Using SNPE}
There are two caveats regarding the theoretical proprieties of SNPE: the lack of smoothness in the posterior of kinked models and the issue of modeling stochastic singularities. Many, but not all, s-S models and other models with kinks, may have discontinuities in the posterior distribution. In particular the assumption that there is a normalizing flow parametrization that can generate the posterior is violated. This problem is not unique to flows, MSM will also have similar problems, but because it's not a full information technique ignoring this information will allow for convergence. That being said, in practice, just like MSM, the flow will still converge. However, it will often learn a continuous function that approximates the discontinuity in the posterior. In the maximum likelihood case, a continuous function can approximate the discontinuity well, as dynamic models often have discontinuities in only a handful of points if that. Thus, the estimation result will generally be near the true likelihood, which is still an efficiency improvement over the MSM. That being said, if estimation with a kinked model is a problem, one can also add measurement error to smooth the problem out.    

Stochastic singularities are the second pathological example that the model will have difficulty handling. This also leads to some estimation issues, not unique to flows, but also other MLE approaches, like maximum likelihood after solving with perturbation. When this happens, the dynamic model learns a sub-manifold on the entire space. Because a normalizing flow has to have positive support on the entire space of of the target distribution, this also violates the assumption that the normalizing flow parameterization can generate the posterior. Like in the previous case, the flow will just learn to put an asymptotically small probability on spaces not in the dynamic models support. That being said, in the cases where there are stochastic singularities and one uses real data that doesn't lie on the sub-manifold of the model, the model will fail to match the data. MSM will typically still match the data, but this a weakness of the MSM. MSM will converge because it cannot discern that the model has no support in those region of the data. One can add more shocks, add measurement error, or a change in modeling like the generalized sS model \citep{caballero2007price} to avoid a stochastic singularity. If the true data is on the submanifold, SNPE will work.  

\section{Results} \label{section_results}
Each model has it's own set of results. The first model is the RBC model where I estimate on simulated data and real data. The second model is toy corporate finance model solved via value function iteration and three different simulation-based approaches are used to estimate this model. The third model is the Lucas asset pricing model solved via projection. The forth model is the \citet{smets2007shocks} model estimated via both MH-MCMC and SNPE. The fifth model is a HANK model solved via Reiter's method and time iteration. Finally the sixth model is a model of bequests which is solved via value function iteration. For more information on the details and construction of the models, refer to the appendix. All six models have been estimated using my simulation-based inference approach.  

I will now discuss proper posterior behavior and how to inspect the accuracy of the posterior through inspection. While inspection is used for all of the models, for some that can be estimated reasonably, I will compare posteriors with MH-MCMC. Through the Bernstein-von Mises theorem, the posterior should converge to the true parameter given enough data samples. However, in finite sample sizes the correct posterior may not match the value of the true parameters. Nevertheless, it is a useful check to make sure that the posterior mode generally matches in at least some of the parameters and doesn't put vanishing posterior mass at the true values. 

\subsection{RBC Model}
The first set of results deal with the RBC model. The RBC model has 4 parameters: $\alpha$, $\beta$, $\delta$, and $\rho$ as defined in the economic models section of the appendix. The CRRA elasticity, $\gamma$ is set to 2. The model is simulated for 200 iterations and the first 100 iterations is dropped to get a steady state behavior. The results of the posterior with the parameters displayed in the same order as above is shown here: \newline
\begin{figure}[H]
    \centering
    \includegraphics[width=9cm]{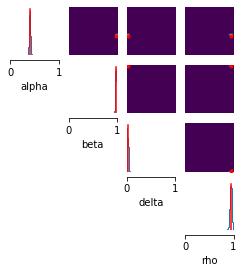}
    \caption{Simulated Data RBC}
\end{figure}
The red bar indicates the true parameter values and the blue line indicates the posterior of the model as estimated by the simulation-based inference approach. Each of the density plots in the upper triangular portion of the graph indicates two way densities given by the corresponding row and column. Likewise the red dot indicates the true value of the parameter with respect to the two axis. In all models, uniform priors are used along the interval specified for all parameters.   

Next I show a chart that displays the an approximate ground truth by running MH-MCMC for 2 million iterations. Although in the limit of time steps, given regularity conditions, the Bernstein-von Mises theorem implies that the Bayesian posterior converges to the maximum likelihood solution (which in this case is the true value given data simulated from a parameterization), with a set number of data points, the posterior may not concentrate around the true parameters. The fact that MH-MCMC concentrates around the true solution, indicates that the posterior obtained with SNPE is likely the true posterior.  

\begin{figure}[H]
    \centering
    \includegraphics[width=9cm]{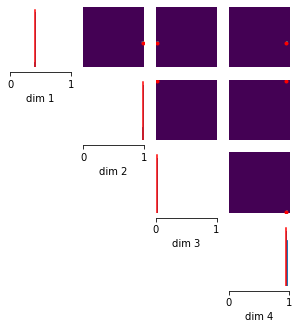}
    \caption{Simulated Data RBC}
\end{figure}

As is clear, for MH-MCMC the simulation-based inference approach concentrates almost entirely on the true parameter value. This extremely close relationship is not entirely replicated on the other models, as the other models have flatter likelihoods and more parameters. In many cases the true posterior is not a delta function in the same way the simulated RBC model is.  

The third chart is the same RBC model estimated on real data:
\begin{figure}[htbp]
    \centering
    \includegraphics[width=10cm]{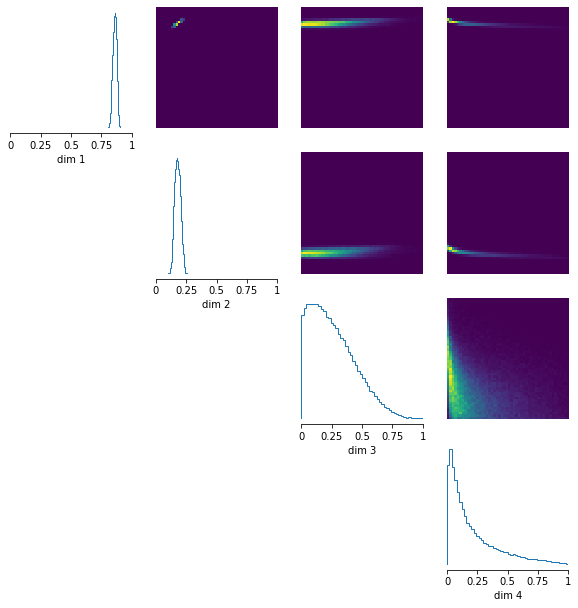}
    \caption{Real Data RBC}
\end{figure}

Here there is no available ground truth but the parameters derived are not realistic, which is not surprising for such a simple model. You can see that compared to the first chart, the method is more uncertain of ground truth. You can also see the model doesn't estimate parameters close to what theory would suggest. This is not surprising given that the RBC model is too simple to work on real data and many of these parameters don't agree with theory even in more complex models. Furthermore, to illustrate the accuracy of the model, all priors are uniform over the charted interval. In the model, it predicts the Cobb-Douglas parameter, $\alpha$ should be very close to 1, which is reasonable since labor is fixed in this model and capital is highly correlated with labor, combined with the fact that a Cobbs-Douglas function is homogeneous with degree 1. The discount rate, $\beta$, also doesn't agree with theory, which is unsurprising as $\beta$ is often calibrated because it's so difficult to pin down. Likewise the parameter on the productivity process, $\rho$, is another variable that is often estimated but is difficult to pin down in this model. $\delta$ or the depreciation rate, is the only parameter where the model is relatively reasonable.  

\subsection{Cash Flow Model}
The next set of results deal with a partial equilibrium model of firm cash flow solved via value function iteration. The parameters here are $\alpha$, $\delta$, $\sigma$, and $\rho$. $\sigma$ is the standard deviation of the productivity law of motion. The interest rate is set at 5 percent and $\beta$ is set as 1/(1+r). The cash flow model was estimated with three different methods, each of which obtaining nearly the same posterior, giving confidence the approach has converged to the right solution. The first chart will show the posterior for a simulation-based inference approach that uses a normalizing flow and a feed forward network as the embedding network (converting the high dimensional $x$ data into a lower dimensional conditioning variable). The second chart will show the same model and data, but using a RNN embedding network. The third chart will show a density estimator that is a GAN instead of a normalizing flow. This is an alternative simulation-based estimator: Sequential Neural Ratio Estimation which is discussed in the appendix.

Displaying the charts:

\begin{figure}[H]
    \centering
    \includegraphics[width=10cm]{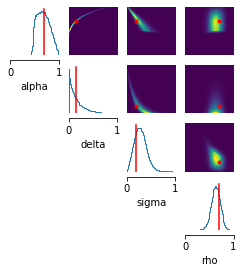}
    \caption{Value Function Iteration with a Dense Net Embedding}
\end{figure}
\begin{figure}[H]
    \centering
    \includegraphics[width=10cm]{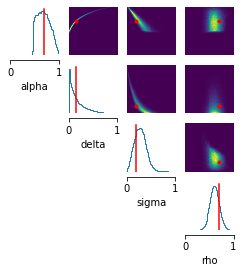}
    \caption{Value Function Iteration with an RNN Embedding}
\end{figure}
\begin{figure}[H]
    \centering
    \includegraphics[width=10cm]{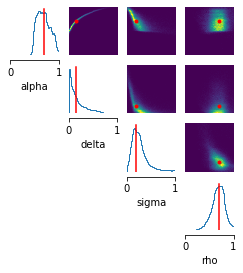}
    \caption{Value Function Iteration with a GAN Density Estimator}
\end{figure}

Despite using different inference techniques, the posteriors look fairly equivalent by visual inspection, suggesting the same distribution has been learned. Furthermore it does seem like the mode is fairly close to the actual parameterization. There some higher order considerations that are getting in the way of the dynamics of the model, but the model seems to be able to accurately recover the true parameters.  

\subsection{Lucas Asset Pricing Projection Model}

The following section shows the estimation procedure on a Lucas asset pricing model:
\begin{figure}[H]
    \centering
    \includegraphics[width=10cm]{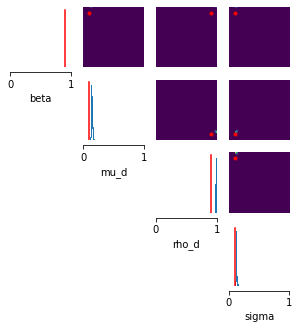}
    \caption{Ten Parameter HANK model}
\end{figure}

The model does seem to accurately fit reproduce the true values. the main error is on $\rho_d$ where the posterior seems to be higher than the true value for the autogregressive parameters.  All other posteriors match very closely the actual data generating parameters. This model took under 10 hours to estimate on an 8 core Intel i7 machine.  

\subsection{Smets-Wouters 2007 Model}
A description of the model can be found at \citet{smets2007shocks}. The model is a representative agent New Keynesian model. The model has 8 shocks that interact with the rest of the model in a autoregressive manner and 8 observed variables.  The model estimates a model of potential GDP with a flexible price economy, along with the true economy with sticky Calvo prices. The model is expressed in linearized form.  For more information on the model see the paper and the appendix.  Since there are 37 parameters in this model, the data are seperated into three different charts of 12,12,13 parameters each.  

\begin{figure}[H]
    \centering
    \includegraphics[width=12cm]{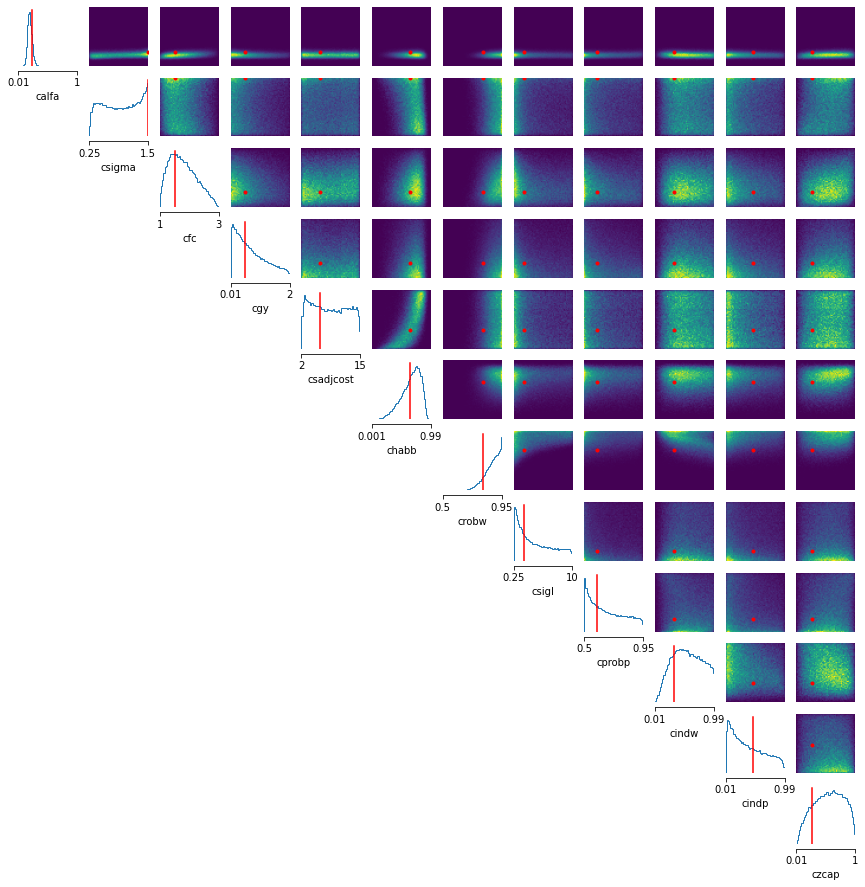}
    \caption{Smets-Wouters Posterior with 1-12}
\end{figure}

\begin{figure}[H]
    \centering
    \includegraphics[width=12cm]{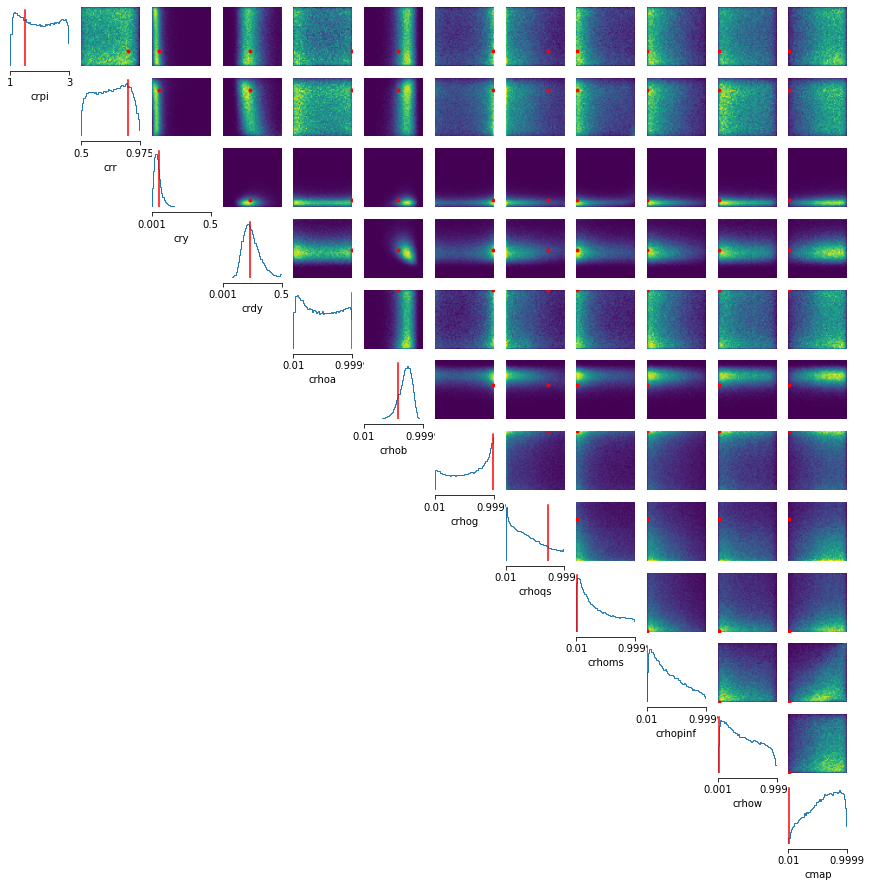}
    \caption{Smets-Wouters Posterior with 12-24}
\end{figure}

\begin{figure}[H]
    \centering
    \includegraphics[width=12cm]{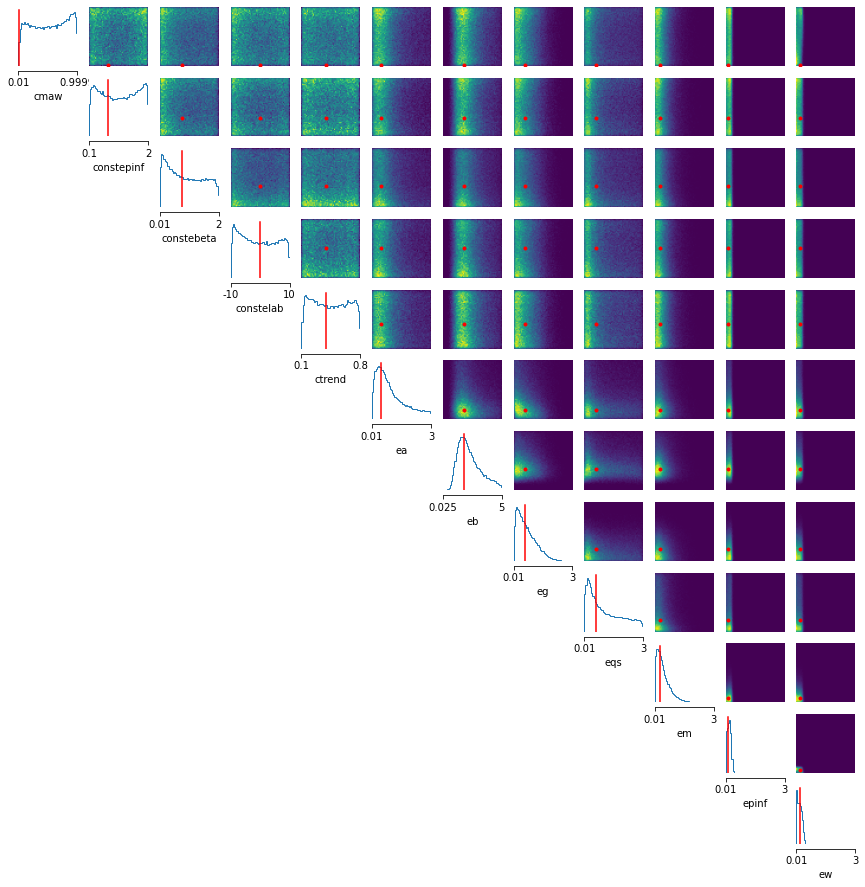}
    \caption{Smets-Wouters Posterior with 24-37}
\end{figure}

I will next discuss some of the parameters and the accuracy of the model. The model successfully estimates capital share (calpha). The model seems to have a difficult time estimating the elasticity of the adjustment costs (csadjcost) as the posterior is diffuse. Fixed costs (cfc) seems to be estimated well, with the mode corresponding to the data generating parameter. The wage stickiness posterior (cprobw) is generally in the right place but there is little probability mass in the location of the true parameter. The wage indexation parameter (cindw) seems to be well identified. The Taylor rule parameters for output (cry) and change in output (crdy) also agree with the ground truth, but the inflation Taylor rule parameter (crpi) does not. Some autoregressive parameters on shocks seem to be well estimated (crhob, crhoms, crhow), while others seem to be diffuse or in areas of low probability mass (crhoa, cmaw, crhoqs, cmap). Typically these parameters are hard to identify, but the model seems perform adequately in identifying some of these parameters. Surprisingly, the model seems to do a poor job in identifying the mean of the observed variables (ctrend, constepinf, conster, constelab, constebeta), but performs well in identifying the standard deviation of shock variables (ea, eb, eg, eqs, em, epinf, ew). 

The algorithm can also find multimodal distributions. You can see this with a few of the parameters, namely crpi and constepinf among others. Since the model estimates posterior by estimating a density from joint distribution draws, there is no reason multi-modal distributions would be more problematic than uni-modal distributions. However with MH-MCMC, multi-modal distributions are a well documented problem \citep{pompe2020framework}, \citep{herbst2015bayesian}. A more elaborate discussion of the model structure which links up all the parameter names to the Smets-Wouters equations is found in the appendix. 

I will now discuss the Smets-Wouters model estimated with MH-MCMC. Although the models were estimated on different computers, each model was solved 1 million times. While estimating the density from samples adds overhead for SNPE, the initial burn in period and the use of the Kalman filter also adds overhead for MH-MCMC. Despite the use of different computers, the amount of overhead is comparable for the two approaches with SNPE using slightly more computation. Below shows the posterior after MH-MCMC:

\begin{figure}[H]
    \centering
    \includegraphics[width=12cm]{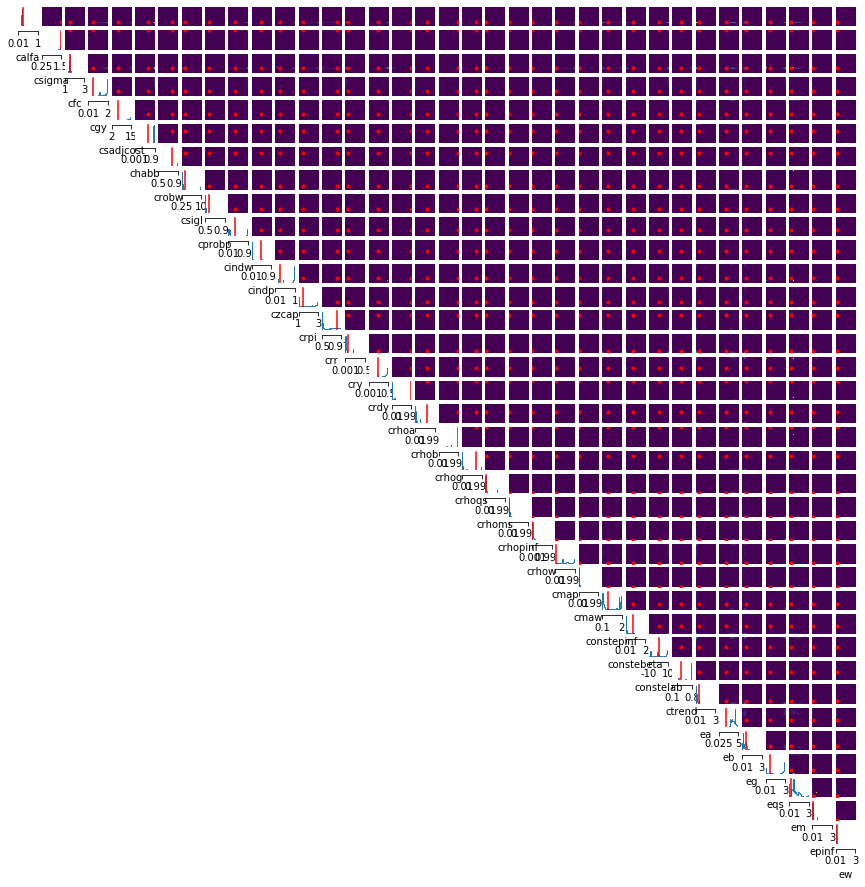}
    \caption{Smets-Wouters Posterior with MH-MCMC}
\end{figure}

As you can see, the quality of the posterior compares poorly to SNPE. Many times the mode is far from the true parameter value. Additionally, quite a few parameters have negligible mass near the true parameter value and the distribution of the posterior looks much more scattered and disconnected than in the SNPE case. Nevertheless checks indicate the MH-MCMC algorithm is sampling from places of high likelihood, it's just not sampling for the entire array of locations where this is the case.  

\subsection{HANK Model}
The fifth model is the 10 parameter HANK model with the parameters discussed in the model section:  
\begin{figure}[H]
    \centering
    \includegraphics[width=10cm]{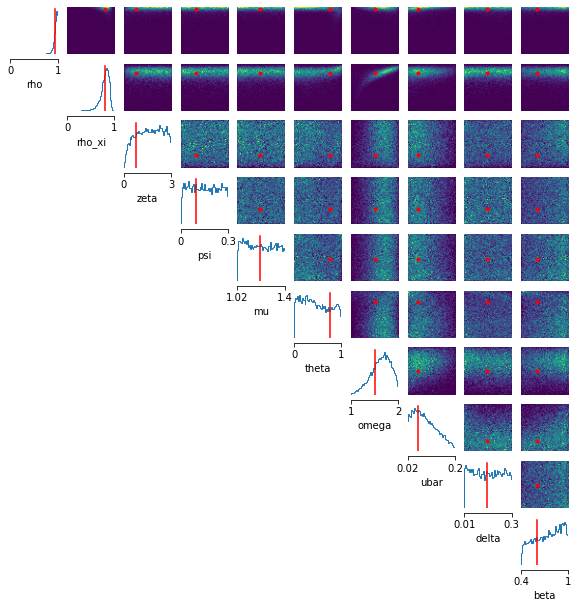}
    \caption{Ten Parameter HANK model--Partially Observed}
\end{figure}

I will next discuss the estimation of the HANK model. This model has only two shocks and thus only two observable equations, the unemployment rate (which is linked to output via a labor only production function) and the interest rate.  Since there were 810 latent variables in the model, it's perhaps reasonable to expect that many of the parameters were difficult to identify. That being said the parameters the approach was confident in aggreed with the ground truth. This illustrates the the model can estimate HANK models with large latent state space--something that is computationally intractable for a likelihood based approach without dimension reduction. For the parameters rho, rho-xi, omega and ubar, the mode of the distribution is either at the true parameter value or in the case of omega, close to the true parameter value. The other parameters all exhibit flat likelihoods, which may not be surprising considering the limited impact of some of the parameters on the two chosen observed variables.



\subsection{Bequests Model}
The sixth model is the 11 parameter bequests model with parameters displayed in the chart corresponding to the parameters in the model appendix section:

\begin{figure}[H]
    \centering
    \includegraphics[width=10cm]{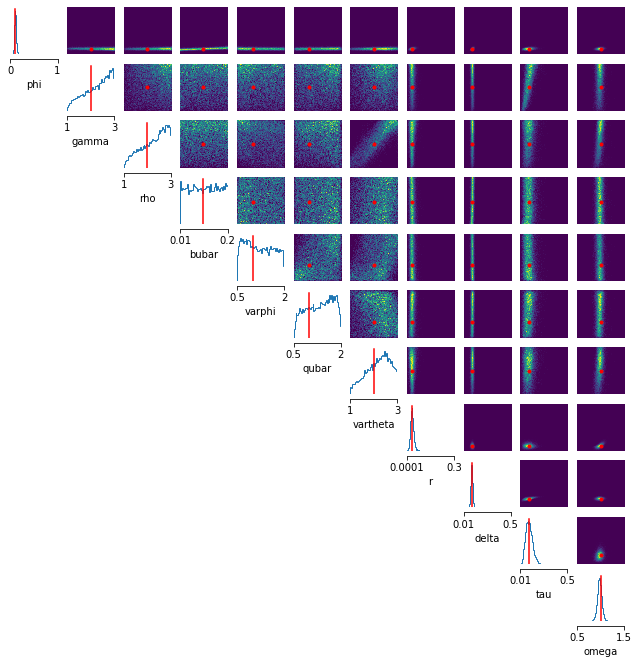}
    \caption{Eleven Parameter Model Estimated via VFI}
\end{figure}

While many of the parameters have converged to the value of the true parameters, some of the parameters still have high uncertainty. This could be due to the fact that the parameters have very flat likelihoods. Regardless, this, the HANK model, and the Smets-Wouters models are all research scale, and the estimation of posteriors seems good in all cases. Many of the posterior distributions do concentrate around the true value, suggesting the model does have some discerning power with regard to the given parameters. Furthermore this is a reasonable sized model solved via value function iteration and the algorithm succeeds relatively well in calculating the posterior.  

\section{Conclusion} \label{section_conclusion}
In this paper, I propose a method to estimate models in a Bayesian or MLE manner using simulations only, without a likelihood function. This works even on models that have a latent state space relationships across time. This model can Bayesian estimate HANK models with large latent state spaces and also can be applied to solution methods like value function iteration and projection which don't yield a likelihood function. The model can be applied as broadly  as MSM, with better gaurantees like efficiency and posterior estimation. Additionally, outside of macroeconomics, this approach can be applied to industrial organization, applied game theory, and econometrics. In industrial organization and game theory, this could be used to solve structural and dynamic models. In econometrics, general latent time series models can be estimated more effectively and efficiently, without likelihood functions. Additionally, the normalizing flow which is a bijective map that can perform arbitrary change of variables is a powerful technique that hasn't been explored in depth. For example, normalizing flows can be applied to the identification in non-separable models literature and can be used as transport maps in optimal transport. Despite the focus on macroeconomic modeling, the application of this technique is more general and has much greater applications that is beyond the scope of this paper.   

\bibliographystyle{aea.bst}
\bibliography{refs.bib}

\appendix

\section{Economic Models}
\label{section_economic_models}
The models used to test these algorithms are disparate. I estimate a relatively standard RBC model via time iteration, a model of capital formation solved via value function iteration, a Lucas Tree model \citep{lucas1978asset} solved with projection, the linearized Smets-Wouters 2007 model, a HANK model solved via Rieter's method with time iteration and lastly a 11 parameter bequests model solved via value function iteration. I will now discuss each of these models sequentially.  

\subsection{The RBC Model}
This RBC model is standard and comes from Alisdair McKay's note on heterogeneous agent models: \url{https://alisdairmckay.com/Notes/HetAgents/index.html}. 

The utility function is a standard CRRA utility:
\begin{equation}
    u(C_t) = E\sum_t\beta^t\dfrac{C_t^{1-\gamma}}{1-\gamma}
\end{equation}

The production function is standard Cobb-Douglas with productivity:
\begin{equation}
    Y_t = Z_t K_t^\alpha\bar{L}^{1-\alpha}
\end{equation}
Where $\bar{L}$ is fixed labor and $Z_t$ is productivity and $K_t$ is capital which evolves according to a standard law of motion:
\begin{equation}
    K_t = (1-\delta)K_{t-1} + Y_t - C_t
\end{equation}
Productivity evolves according to a AR(1) process in logs:
\begin{equation}
    log Z_t = \rho log Z_{t-1} + \epsilon_t
\end{equation}

This model is solved via time iteration and matches consumption, investment, productivity to real data, interest rates and capital are the latent unobserved variables. This is done on data both synthetic and real.  

\subsection{The Model of Capital Formation}
This model is a straightforward infinite horizon partial equilibrium model with risk neutral managers choosing investment to maximize the present value of a stream of firm cash flows and comes from solution code written by Emil Lakkis and Toni Whited.

Utility/cash flows are give by this formula:
\begin{equation}
    U(K_t,I_t,Z_t) = \sum_t \beta^t*(\pi(K_t,Z_t)-I_t)
\end{equation}
Where $K_t$, $Z_t$, and $I_t$ are the same as in the RBC model--capital, productivity and investment respectively.  $\pi$ is the profit function.  

The capital law of motion is like the RBC model:
\begin{equation}
    K_t = (1-\delta)K_{t-1} + I_t
\end{equation}

Productivity evolves according to a AR(1) process in logs:
\begin{equation}
    log Z_t = \rho log Z_{t-1} + \epsilon_t
\end{equation}

Profit is defined as:
\begin{equation}
    \pi(K_t,Z_t) = Z_t K_t^{\alpha}
\end{equation}

This model shares similarities with the RBC model, with the main difference being the utility function and the value function solution method. This estimation method matches capital levels generated by a parameterized model. 

\subsection{Lucas Asset Pricing Projection Model}
Each agent maximizes CRRA utility:
$$
    max_i E_t[\beta^t\dfrac{c_t^{1-\gamma}-1}{1-\gamma}
$$

with $\gamma$ set to 3. The law of motion is:
$$
    k_{t+1} = (1+d_t/P_t)k_t - c_t/P_t
$$

The state of the system is defined as dividends received plus price of the amount of assets owned:
$$
    m_{t+1} = (P_{t+1}+d_{t+1})k_{t+1}
$$

Putting these three equations together one gets the pricing kernel:

$$
    p_t = \beta^t E_t[(\dfrac{c_{t+1}}{c_t})^{-\gamma}(d_{t+1}+p_{t+1})]
$$

Dividends evolve under an $AR(1)$ process:

$$
d_{t+1} = \mu_d + \rho_d*d_t + \epsilon
$$

Where $\epsilon$ is a shock with standard deviation $\sigma$. When behaving optimally, consumption also equals the dividends received. This model matches the expected return and risk free rate at a variety of dividend levels, which demonstrates the ability of this model to not only estimate on models solved via projection, but also to work on cross sectional problems as well.  

\subsection{Smets-Wouters 2007 Model}
This section will mainly illustrate the equations and the parameters so one can cross reference parameters shown in the diagram with the correct equation in the DSGE model. This model comes from the paper \citet{smets2007shocks}. More details on the economics of the equations can be found there. 

First, a brief discussion of the overview of the model. The model is a representative New Keynesian model with stickiness in both wages and prices. The model has a flexible price set of equations which determines potential GDP for monetary policy and a sticky price system with a price markup equation that determines the true economy. I will only be discussing the sticky price system of equations with an implicit understanding that there is a mirroring flexible price system. Shocks all have autoregressive components to more accurately match data. All equations are linearized. 

The aggregate resource constraint is:
\begin{equation}
    y_t = ccy*c_t+ciy*i_t+\epsilon^g_t  +  1*crkky*z_t
\end{equation}

The Euler equation is given by: 
\begin{equation}
\begin{aligned}
    c_t = &(chabb/cgamma)/(1+chabb/cgamma)*c_{t-1} + (1/(1+chabb/cgamma))*c_{t+1} + \\ &((csigma-1)*cwhlc/(csigma*(1+chabb/cgamma)))*(l_t-l_{t+1}) - \\ &(1-chabb/cgamma)/(csigma*(1+chabb/cgamma))*(r_t-\pi_{t+1}) +\epsilon^b_t
\end{aligned}
\end{equation}

The investment Euler equations are:
\begin{equation}
    i_t = (1/(1+cbetabar*cgamma))* (i_{t-1} + cbetabar*cgamma*i_{t+1}+(1/(cgamma^2*csadjcost))*q_t) + \epsilon^i_t
\end{equation}

The arbitrage equation for the value of the capital stock is:
\begin{equation}
\begin{aligned}
     q_t =&-r_t+\pi_{t+1} +(1/((1-chabb/cgamma)/(csigma*(1+chabb/cgamma))))*\epsilon^b_t + \\ &(crk/(crk+(1-ctou)))*r^k_{t+1} +  ((1-ctou)/(crk+(1-ctou)))*q_{t+1}
\end{aligned}
\end{equation}

The linearized production function is:
\begin{equation}
     y_t = cfc*(calfa*k_t+(1-calfa)*l_t +\epsilon^a_t)
\end{equation}

Capital utilization:
\begin{equation}
    k^s_t =  k_{t-1}+z_t
\end{equation}

Capital utilization is some function of the interest rate for capital:
\begin{equation}
    z_t = (1/(czcap/(1-czcap)))*r^k_t
\end{equation}

Capital's law of motion is:
\begin{equation}
    k_t =  (1-cikbar)*k_{t-1}+cikbar*i_t + cikbar*cgamma^2*csadjcost*\epsilon^i_t
\end{equation}

The markup equation is:
\begin{equation}
    \mu^p_t = calfa*k^s_t+(1-calfa)*w_t - \epsilon^a_t 
\end{equation}
This equation is the only set of economic equations that doesn't have a flexible price counterpart. Note that the observed variable equations also will not have a flexible price counterpart.  

The New Keynesian Phillips curve:
\begin{equation}
\begin{aligned}
    \pi_t =&  (1/(1+cbetabar*cgamma*cindp))*(cbetabar*cgamma*\pi_{t+1} + cindp*\pi_{t-1} +\\ 
    &((1-cprobp)*(1-cbetabar*cgamma*cprobp)/cprobp)/((cfc-1)*curvp+1)*(\mu^p_t)  )  + \\ 
    &\epsilon^p_t
\end{aligned}
\end{equation}

The rental rate of capital equation:
\begin{equation}
    r^k_t =  w_t+l_t-k_t
\end{equation}

The path of wages is:
\begin{equation}
\begin{aligned}
     w_t =&(1/(1+cbetabar*cgamma))*w_{t-1}+(cbetabar*cgamma/(1+cbetabar*cgamma))*w_{t+1}+ \\
        &(cindw/(1+cbetabar*cgamma))*\pi_{t-1}-(1+cbetabar*cgamma*cindw)/(1+cbetabar*cgamma)*\pi+ \\
        &(cbetabar*cgamma)/(1+cbetabar*cgamma)*\pi_{t+1} + \\
        &(1-cprobw)*(1-cbetabar*cgamma*cprobw)/((1+cbetabar*cgamma)*cprobw)*\\
        &(1/((clandaw-1)*curvw+1))*(csigl*l_t + (1/(1-chabb/cgamma))*c_t - \\ &((chabb/cgamma)/(1-chabb/cgamma))*c_{t-1} -w_t) + \epsilon^w_t   
\end{aligned}
\end{equation}

The Taylor rule followed by central banks is:
\begin{equation}
     r_t =  crpi*(1-crr)*\pi_t+cry*(1-crr)*(y_t-y^p_t)+crdy*(y_t-y^p_t-y_{t-1}+y^p_{t-1})+crr*r_{t-1}+\epsilon^r_t 
\end{equation}

The observed variables equations are:
\begin{align}
    &dy_t=y_t-y_{t-1}+ctrend\\
    &dc_t=c_t-c_{t-1}+ctrend\\
    &di_t=i_t-i_{t-1}+ctrend\\
    &dw_t=w_t-w_{t-1}+ctrend\\
    &\pi^o_t = \pi_t + constepinf\\
    &r^o_t = r_t + conster\\
    &l^o_t = l_t + constelab
\end{align}

The autoregressive equations for shocks are:
\begin{align}
    &\epsilon^a_t = crhoa*\epsilon^a_{t-1}  + ea\\
    &\epsilon^b_t = crhob*\epsilon^b_{t-1} + eb\\
    &\epsilon^g_t = crhog*\epsilon^g_{t-1} + eg + cgy*ea\\
    &\epsilon^i_t = crhoqs*\epsilon^i_{t-1} + eqs\\
    &\epsilon^r_t = crhoms*\epsilon^r_{t-1} + em\\
    &\epsilon^p_t = crhopinf*\epsilon^p_{t-1} + epinfma - cmap*epinfma(-1)\\
    &epinfma=epinf\\
    &\epsilon^w_t = crhow*\epsilon^w_{t-1} + ewma - cmaw*ewma(-1) \\
    &ewma=ew
\end{align}

Some of the higher level parameters in the above equations are also related to the parameters in my diagram by the following equations:

\begin{align}
    &cpie = 1 + constepinf/100 \\
    &cgamma = 1 + ctrend/100\\
    &cbeta = 1/(1+constebeta/100)\\
    &clandap=cfc\\
    &cbetabar=cbeta*cgamma^{(-csigma)}\\
    &cr=cpie/(cbeta*cgamma^{(-csigma)})\\
    &crk=(cbeta^{(-1)})*(cgamma^{csigma}) - (1-ctou)\\
    &cw = \dfrac{(calfa^{calfa}*(1-calfa)^{(1-calfa)}}{(clandap*crk^{calfa}))^{(1/(1-calfa))}}\\
    &cikbar=(1-(1-ctou)/cgamma)\\
    &cik=(1-(1-ctou)/cgamma)*cgamma\\
    &clk=((1-calfa)/calfa)*(crk/cw)\\
    &cky=cfc*(clk)^{(calfa-1)}\\
    &ciy=cik*cky\\
    &ccy=1-cg-cik*cky\\
    &crkky=crk*cky\\
    &cwhlc=(1/clandaw)*(1-calfa)/calfa*crk*cky/ccy\\
    &cwly=1-crk*cky\\
    &conster=(cr-1)*100
\end{align}

The seven observed equations $dy_t$, $dc_t$, $di_t$, $dw_t$, $\pi^o_t$, $r^o_t$, $l^o_t$, are matched using data simulated from a model with true parameters equal to the Smets-Wouters 2007 mean parameters, with slight modifications as some parameters were calibrated to 0.      

\subsection{The HANK Model}
The HANK model also comes from Alisdair McKay's tutorial, which adds a heterogeneous distribution of wealth as a state variable for households combined with a New Keynesian backbone. 

Preferences are still CRRA, like the RBC model. The firm side is now new Keynesian with intermediate goods produced via:
\begin{equation}
    y_{j,t} = Z_t N_{j,t}
\end{equation}
Here $Z_t$ is aggregate productivity and $N_t$ is labor employed by firm j producing variety $y_{j,t}$.  The final good is produced via the Dixit-Stiglitz aggregator \citet{dixit1977monopolistic}:
\begin{equation}
    Y_t = (\int_0^1y_{j,t}^{\dfrac{\epsilon - 1}{\epsilon}})^{\dfrac{\epsilon}{\epsilon - 1}}
\end{equation}
Here $\epsilon$ is the elasticity of substitution between intermediates.  

Output is defined as,

$\begin{aligned}
Y_t = A_t \int_0^1 n_{j,t} dj = A_t N_t
\end{aligned}$

Where $n_{j,t}$ is the labor for the individual intermediate goods. Additionally from the accounting identity, output is also,
$\begin{aligned}
Y_t = C_t + \psi M_t H_t
\end{aligned}$

Where the second term is hiring costs.

Government and bonds are defined by the following nominal and real budget constraint equations:
\begin{align}
    &\tau P_t (w_t+d_t) (1-u_t) + P_t B = (1+i_{t-1}) P_{t-1} B + P_t b u_t \\
    &\tau (w_t+d_t) (1-u_t) + B = R_{t} B + b u_t
\end{align}
Labor market is a standard McCall Search model:

\begin{align}
&N_t = (1-\delta) N_{t-1} + H_t \\
&1-u_t = (1-\delta) (1-u_{t-1}) + H_t \\
&M_t = \frac{ H_t }{u_{t-1} + \delta N_{t-1}} \\
&M_t = \frac{ 1-u_t  - (1-\delta) (1-u_{t-1})}{ u_{t-1} + \delta (1-u_{t-1})}
\end{align}

The household maximizes the CRRA utility like the RBC model. The households budget constraint in real money is:
\begin{equation}
    a_t + C_t = R_t*a_{t-1} + earnings
\end{equation}
Here $a_t$ is savings and $R_t$ is the interest rate as determined by the bond market clearing condition. Earnings are given by:
\begin{equation}
    \begin{split}\begin{aligned}
    \mbox{earnings} =  \begin{cases} (1-\tau_t) \left(w_t + d_t \right) & \mbox{if employed} \\
    b & \mbox{if unemployed,}
    \end{cases}\end{aligned}\end{split}
\end{equation}

Here $w_t$ is wages, $d_t$ is dividend income and $b$ is unemployment benefits. 

The stochastic matrix governing employment, unemployment transitions is given by:
\begin{equation}
    \begin{split}\left( \begin{matrix} 1-M_t & \delta \\ M_t &1-\delta \end{matrix} \right)\end{split}
\end{equation}

Again $M_t$ is governed by the labor market equations above.  

The firm solves the Dixit-Stiglitz cost problem:
\begin{equation}
    y_{j,t} = \left( \frac{p_{j,t}}{ P_t} \right)^{-{\varepsilon}} Y_t
\end{equation}
with price index $P_t$ given by
\begin{equation}
    P_t^{1-{\varepsilon}} = \int_0^1 p_{j,t}^{1-{\varepsilon}} dj.
\end{equation}

The Calvo pricing adjustment parameter $\theta$ implies that the intermediate firm chooses price $p*$ to maximize:
\begin{equation}
    \mathbb E_t \sum_{s=t}^\infty \theta^{s-t} R_{t,s}^{-1} \left[ \frac{p_t^*}{P_s} y_{j,s} - w_s n_{j,s} - \psi M_s h_{j,s}\right]
\end{equation}
Here $n_{j,t}$ denotes employment at firm $j$ in time $t$. $h_{j,t}$ denotes hiring and $R^{-1}_{t,s}$ denotes the real interest rate discounted $s$ periods back to time $t$. This model is solved using Reiter's method in combination with time iteration. This model matches unemployment and interest rate to the same data generated by a parameterized model, only two observed variables are used as the model only had two shocks.  

\subsection{The Value Function Iteration Bequests Model}
Finally this is a model of bequests over time. The model illustrates the benefit of my approach as it cannot be estimated by perturbation because of the kink at the intersection of the consumption versus adjusting illiquid savings intersection. Thus one has to use simulation-based inference to estimate this model solved via value function iteration.  

This model comes from Jeppe Druedahl's open source model example: \url{https://github.com/pkofod/vfi/blob/master/Fast\%20VFI.ipynb}. The utility function is:
\begin{eqnarray}
u(b_{t},c_{t}) = \frac{[\phi(b_{t}+\underline{b})^{1-\gamma}+(1-\phi)c_{t}^{1-\gamma}]^{\frac{1-\rho}{1-\gamma}}}{1-\rho}
\end{eqnarray}

Here $b_t$ is bequest or non-liquid savings.  $c_t$ is consummation.  $b, \phi, \gamma, \rho$ are all parameters. The bequest utility function (end of life utility) is:
\begin{eqnarray}
\nu(a_{t},b_{t})=\varphi\frac{(a_{t}+b_{t}+\underline{q})^{1-\vartheta}}{1-\vartheta}
\end{eqnarray}

Here $a_t$ is liquid savings and $q, \theta, \varphi$ are all parameters. The value function is give by:
\begin{eqnarray}
v_{t}(m_{t},n_{t},l_{t})&=&\max\{v_{t}^{keep}(m_{t},n_{t},l_{t}),v_{t}^{adj}(m_{t},n_{t},l_{t})\} \\
& \text{s.t.} & \\ 
x_{t}&=&m_{t}+(1-\tau)n_{t}
\end{eqnarray}

Here $m_t$ is beginning of period liquid savings (analogous to end of period $a_t$) and $n_t$ is beginning of period illiquid savings (analogous to end of period $b_t$).  $x_t$ is the pooled value of total liquid and illiquid savings. $l_t$ is labor income.  The post-decision value function is: 
\begin{eqnarray}
w_{t}(a_{t},b_{t},l_{t}) &=& \nu(a_{t},b_{t}), t = T \\
w_{t}(a_{t},b_{t},l_{t}) &=& \mathbb{E}_{t}\left[\max\{v_{t}^{keep}(m_{t+1},n_{t+1},l_{t+1}),v_{t}^{adj}(x_{t+1},l_{t+1})\}\right], t < T \\
&\text{s.t.}& \\
l_{t+1}&\sim& F(l_{t}) \\
m_{t+1}&=& (1+r_{a})a_{t}+\omega l_{t+1} \\
n_{t+1}&=& (1-\delta)b_{t} \\
x_{t+1}&=& m_{t+1}+(1-\tau)n_{t+1}
\end{eqnarray}
Here labor income comes from a distribution $F(l_t)$.  Likewise $r_a, \omega, \delta, \tau$ are all parameters. The keep value function is: 
\begin{eqnarray}
v_{t}^{keep}(m_{t},n_{t},l_{t}) &=& \max_{c_{t}\in[0,m_{t}]}u(n_{t},c_{t})+\beta w_{t}(a_{t},b_{t},l_{t}) \\
&\text{s.t.}& \\ 
a_{t}&=& m_{t}-c_{t}\\b_{t}&=& n_{t}
\end{eqnarray}
The keep value function is the value function at a point $t$ if the agent chooses to consume and not withdraw or deposit funds into the illiquid asset, $b_t$. The adjust value function is:
\begin{eqnarray}
v_{t}^{adj}(x_{t},l_{t}) &=& \max_{b_{t}\in[0,x_{t}]}v_{t}^{keep}(m_{t},n_{t},l_{t}) \\ 
&\text{s.t.}& \\ 
m_{t}&=& x_{t}-b_{t}\\n_{t}&=& b_{t}
\end{eqnarray}
The adjust value function is the value of adjusting the illiquid asset. Notice the agent is unable to consume this period. This model with it's nonlinear and kinked decision rule cannot be effectively modeled with a method like perturbation. Even so, I use my estimation technique to get a Bayesian posterior distribution. This model matches the liquid asset, illiquid asset and consumption to synthetic data generated by a model.

\section{Background on Simulation Neural Ratio Estimators}
\subsection{General Approach of simulation-based Inference}
The differences between the two simulation-based inference approaches roughly centers around what conditional density estimator to use. Sequential Neural Ration Estimators (SNRE) uses a GAN and Sequential Neural Posterior Estimators (SNPE) use a normalizing flow. While one can sample a lot of points from the joint, an iterative approach where one fits the density estimator on sampled data then re-samples leads to lower variance. I will start discussing SNRE, by introducing the GAN

\subsection{Generative Adversarial Networks}
\label{GAN}
The GAN is an algorithm that uses two competing models moving in tandem to generate data that matches data generated from the real world \citep{goodfellow2014generative}. Below is a diagram describing the structure of a basic GAN:

\begin{figure}[htbp]
    \centering
    \includegraphics[width=10cm]{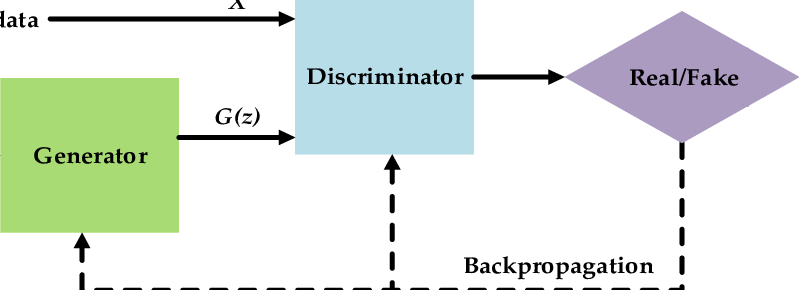}
    \caption{Diagram of a GAN}
    \label{fig:GAN diagram}
\end{figure}

The generator is a neural network (or any differentiable model like a dynamic macro model) which attempts to produce data (unconditionally or conditionally via some inputs $Z$) that resembles data in the real world. The discriminator is another neural network (or any trainable model that can produce a logistic output), which outputs the probability a given input data comes from the underlying data generating process and not from the generator. The two networks play a game where the objective of the generator is to convince the discriminator to assign high likelihood to its output and the objective of the discriminator is to assign the correct probabilistic label, a number between 0 or 1, to input that is either generated data or real data. One way to think about a GAN as a simulated method of moments algorithm where the moment conditions automatically evolve over many rounds to pick the most mismatched moments between the true data and the model. When thought in this way, it's intuitive to recognize that the GAN is nearly a full information and efficient estimation procedure. For more thorough proofs see \citet{kaji2020adversarial}. 

More mathematically, the objective of the discriminator is to minimize a logit loss of input that comes either from real data $x$, or the generator $G(z)$ where the $z$ are shocks that generate stochasticity in the GANs generation. Mathematically the discriminator loss is:

\begin{align}
Loss_D = -E_{x\sim P_{data}(x)}[D(x)] - E_{z\sim P_{sim}(z)}[1-D(G(z))]  \label{eqn:GAN1}
\end{align}

This is just a logit loss applied to data either coming from the data generating process or the generator. Likewise the generator's loss is to fool the discriminator:

\begin{align}
Loss_G =  E_{z\sim P_{sim}(z)}[1-D(G(z))]  \label{eqn:GAN2}
\end{align}

These two loss functions can be compactly represented as:
\begin{align}
G,D = \min_G \max_D E_{x\sim P_{data}(x)}[D(x)] + E_{z\sim P_{sim}(z)}[1-D(G(z))]  \label{eqn:GAN3}
\end{align}

Equation \ref{eqn:GAN3} makes it clear the GAN objective is a minimax game with a solution that is a Nash equlibrium. However what is important for simulation-based estimation is to recover likelihoods/probabilities. The next proposition following \citet{goodfellow2014generative} will show how to do that:\newline

\textbf{Proposition 1.}: For a fixed generator, the optimal discriminator will produce a logistic output:

\begin{align}
D^*(x) =  \dfrac{p_{data}(x)}{p_{data}(x)+p_{generator}(x)} \label{eqn:GAN4}
\end{align}

\textbf{Proof:} Using equation \ref{eqn:GAN1}, for the loss of the generator:
\begin{align}
Loss_D &= \int_x log(D(x))p_{data}(x)dx + \int_z log(1-D(G(z)))p_{generator}(z)dx  \\
    &= \int_x (log(D(x))p_{data}(x) + log(1-D(x))p_{generator}(x))dx  \\
\end{align}

Any discriminator that maximizes the integral: $\int_x (log(D(x))p_{data}(x) + log(1-D(x))p_{generator}(x))dx$, maximizes the integral pointwise over all $x$.  Thus the optimal discriminator will set the first order condition of the intergrand to zero: 

\begin{align}
0 =  \dfrac{p_{data}(x)}{D^*(x)} + \dfrac{p_{generator}(x)}{1-D^*(x)} \label{eqn:GAN5}
\end{align}

Rearranging equation \ref{eqn:GAN5} gets the result of the proposition. $\blacksquare$

Then one can use this proposition along with the discriminator output to estimate a likelihood ratio. 

\subsection{Sequential Neural Ratio Estimation}
\label{SNRE}
In SNRE, \citet{durkan2020contrastive} use a GAN, not to match a underlying data like in proposition 1, but to contrast between two different distributions to estimate a density. In macro, the dataset is typically a single data point with correlated timesteps. Using a GAN to discriminate between real data and fake data would not work because the GAN would learn a delta function for the real data. In order to ameliorate this, the authors use contrastive learning where they differentiate between the generated distribution and the independent distribution, $x,\theta \sim p(x)p(\theta)$, where x is sampled uniformly from the empirical distribution of simulated data. Then they contrast the independent distribution, with the true joint: $x, \theta \sim p(x,\theta) = p(x|\theta)p(\theta)$. Furthermore, knowing the optimal discriminator gives the ratio discussed in \ref{eqn:GAN4}. The authors then recover the ratio: $r(x,\theta) = \dfrac{p(x,\theta)}{p(x)p(\theta)}$, using the fact that $\dfrac{p(x,\theta}{p(x,\theta) + p(x)p(\theta)}$ equals $\dfrac{r}{r+1}$. Thus, the density of the posterior $p(\theta|x) = r(x,\theta)p(\theta)$, where $p(\theta)$ is the prior. This extends \citet{kaji2020adversarial} GAN approach in allowing estimation of models with a latent time dimension. One can extend this to multi-round inference where one uses the GAN as a better proposal distribution than the prior, by sampling $\theta$ from the GAN using MCMC. Then one uses importance sampling to deal with the mismatch in proposal distribution and prior. See \citet{durkan2020contrastive}, \citet{hermans2020likelihood} and the SNRE theoretical sections of the appendix for more information.  

\begin{algorithm}[H]
\SetAlgoLined
\textbf{Input:} Simulator $p(x|\theta)$, prior $p(\theta)$, data $x_0$, discriminator $d_\phi(x,\theta)$, Rounds R, Samples S\;
\textbf{Initialize:} Posterior $p^{(0)}= p(\theta)$, data set D = \{\}\;
\For{$i\gets1$ \KwTo $R$}
{
    Sample $\theta^{(n)}\sim p^{(i-1)}$ for $n = 1...S$ with MCMC\;
    Simulate $x^{(n)} \sim P(x|\theta^{(n)})$ for $n = 1...S$\;
    Concatenate data $D = D \cup \{x^{(n)},\theta^{(n)}\}^S_{n=1}$\;
    \While{$d_\phi(x,\theta)$ not converged}{
        Sample $\{x^{(i)},\theta^{(i)}\}^B_i \sim D$ from D\;
        Sample K contrasting samples for each i sample $\{x'^{(k)}\}^{B*K}_k,\{\theta'^{(k)}\}^{B*K}_k \sim D\{x\},D\{\theta\}$, breaking the joint distribution of $D$ into independent marginals\;
        Train the GAN using the GAN loss function from the two distributions $x,\theta$ and $x',\theta'$ and a softmax/multinomial logit objective $\sum_i^B log(\dfrac{exp(d_\phi(x^{(i)},\theta^{(i)})}{\sum_k^K exp(d_\phi(x^{(k)},\theta^{(k)})})$\;
    }
    Update posterior $p^{(i)}\propto exp(d_\phi(x,\theta))p(\theta)$\;
}
 \caption{SNRE Algorithm}
\end{algorithm}

\section{Proofs}
\subsection{Outline}
There are two main parts of this section: 1.) How the Bayesian estimation procedure works, 2.) The estimation procedure with the normalizing flow density estimator.  

\subsection{Bayesian Estimation}
The Bayesian estimation procedure assumes the existence of a suitable density estimator that has all the properties of the normalizing flow. It will be convenient to assume the properties of the normalizing flow as stated which will be proved in the normalizing flow section (following \citet{huang2018neural}).  

The first step is to draw samples from the joint distribution and I will show this is possible.  Since $P(x,\theta) = P(x|\theta)P(\theta)$, the best procedure is to draw $\theta$ from the prior, then draw $x$ from the likelihood $P(x|\theta)$.  This draws one sample $(x, \theta)$ from the joint distribution.  Since the posterior is defined as $P(\theta|x)$, if one can fit a conditional density estimator on $P(\theta|x)$ on the joint samples $(x,\theta)$ one can get the posterior distribution of $\theta$ by conditioning with the true data, $X$.

This procedure is enough to generate samples from the posterior distributions. However, this procedure has high variance if the prior is far away from the posterior as you are sampling $\theta$ in places where the posterior has little mass. A more computationally efficient way is to sample in multiple rounds.  In the first round one uses the prior distribution, however in later rounds one samples $P(\theta)$ from a proposal distribution which should be the most updated guess on the posterior (denoted with a hat) $\hat{P}(\theta|x)$, then one adjusts via importance sampling for the fact that $\hat{P}(\theta|x)$ is not the same as $P(\theta)$.  \newline

\textbf{Theorem 5} \textit{In multi-round simulation-based inference, estimating from a proposal distribution rather than a prior requires adding an adjustment factor to the density estimation of samples produced. Assuming the flow, $f(\theta|x)$, can converge to any probability distribution in distribution  (Theorem 9), then when one fits $f(\theta|x)*\dfrac{\hat{P}(\theta)}{P(\theta)}$ to the data sampled from the proposal distribution $\hat{P}(\theta)$, one will converge to the true posterior $P(x|\theta)$}\newline

\textbf{Proof}:
In order to perform the adjustments note that $P(\theta|x) = P(x|\theta)P(\theta)/P(x)$.  Call the incorrect proposal distribution $\hat{P}(\theta)$ and define $\hat{P}(\theta|x) = P(x|\theta)\hat{P}(\theta)/P(x)$ which is the conditional distribution that would be obtained by naively training a conditional estimator $f(\theta|x)$ on the joint that contains the incorrect proposal distribution. Then 
\begin{equation}\label{IS}
    \hat{P}(\theta|x) = (P(x|\theta)P(\theta)/P(x))*\dfrac{\hat{P}(\theta)}{P(\theta)} = P(\theta|x)*\dfrac{\hat{P}(\theta)}{P(\theta)}
\end{equation}

Since the joint distribution with the differing proposal distribution has a pdf of $P(x|\theta)\hat{P}(\theta)$ so the conditional distribution which one is drawing samples from has pdf $\hat{P}(\theta|x)$. If one is fitting a density estimator $f(\theta|x)$ on the differing proposal joint data via maximum likelihood as one does with a normalizing flow, one will recover estimates the distribution $\hat{P}(\theta|x)$. However, if one fits the density estimator $f(\theta|x) *(\hat{P}(\theta)/P(\theta))$ (ie adjusting the likelihood by the importance weights), then $f_n(\theta|x)$ converges to $\hat{P}(\theta|x)*(P(\theta)/hat{P}(\theta))$ by the dominated convergence theorem. $\hat{P}(\theta|x)*(P(\theta)/hat{P}(\theta)) = P(\theta|x)$ by eq \ref{IS}. $\blacksquare$  \newline

In multi-round simulation-based inference, the prior is the first proposal distribution $\hat{P}(\theta) = P(\theta)$ and the normalizing flow is the resulting first round estimate of the posterior: $f_1(\theta|x = X)$.  Then for the second round $\hat{P}(\theta) = f_1(\theta|x=X)$, where the proposal is now the estimated posterior recovered in the first round evaluated at the real data. Later rounds iteratively replace the proposal $\hat{P}(\theta) = f_{n-1}$ with the estimated posterior recovered from the previous round (and importance sample).  Because importance sampling requires the ratio: ($\hat{P}(\theta)/P(\theta)$), f needs to be both easily sampled from as well as having easily calculable pdf. A normalizing flow can do this, as illustrated in the below section. 

\subsection{Normalizing Flows in Depth}
A normalizing flow is a complex model.  I will discuss the constituent parts.  Here is a decent description of them on an intuitive level: \url{http://akosiorek.github.io/ml/2018/04/03/norm_flows.html}.  I encourage reading the blog post to understand flows at an intuitive level. This appendix will be more technical and so I would encourage consulting other sources for more intuition.  

The point of a normalizing flow is to estimate the density from samples. The idea is to learn a mapping from an easy to model distribution like the multivariate normal distribution, to the target distribution which is learned via samples. The way this is done is to sample points from this target distribution and take the inverse mapping and get the respective base space points. Since the density of the target space distribution is known using a change of variable formula, one can calculate the density (or likelihood) the flow assigns to any sample point in the target space.  Then one optimizes the parameters of the flow via maximize likelihood.  

Before starting I will prove a couple lemmas: \newline

\textbf{Lemma 1}: \textit{Given a normalizing flow comprised of bijectors stacked one on top of each other. The formula of each bijector:}

$$y^j_i = w(y^i_1,...y^i_{i-1})*\sigma(\alpha(y^i_1,...y^i_{i-1})*y^i_i + \beta(y^i_1,...y^i_{i-1}))$$

\textit{with $\alpha$, $\beta$, and $w$ being outputs from neural networks with all positive weights, monotonic nonlinearities, $\sigma$, as well as all elements from the base distribution $y^i$ are positive. This results in a monotonically increasing bijector in $y^i_n$ for each $n$}\newline

\textbf{Note}: The last requirement that the base distribution is all positive is without loss of generally. One can make all base distributions positive then provide a final monotonic non-linearity that maps the positive half real line to the whole real line like a log transform. \newline

\textbf{Proof}:
To prove that a bijector parametrized in this way is monotonic, take a derivative of $y^j_i$ with respect to $y^i_i$.  
$$
dy^j_n/dy^i_n = d\sigma(a)/da*da/dy^i_n
$$

Where $a$ corresponds to the equation inside of $\sigma$.  Since $\sigma(.)$ is monotonic, $d\sigma(a)/da$ has a derivative greater than 0.  Likewise $da/dy^i_i = \alpha(y^i_1,...y^i_{i-1})$ which by assumption is set to be positive with a correct composition of weight and non negative domain restrictions. Thus each dimension of the bijector is monotonic, making the entire bijector monotonic (increasing). A composition and sum of monotonically increasing functions is monotonically increasing so the entire normalizing flow is monotonically increasing, by induction.  $\blacksquare$ \newline

\textbf{Lemma 2}: \textit{If a set $E$ in $X$ has zero measure and a function $g(X)$ is Lipschitz, then $g(E)$ will also have zero measure}\newline

\textbf{Proof:}
Suppose $E \subset A$ has measure zero.  Then cover $A$ with a countably finite union of increasing compact sets in  $R^n$  such that:
$$ 
A\subset \bigcup\limits_{k=1}^\infty C_k. 
$$

Let $E_k$ denote the intersection of $E$ with $C_k$. Define a measure on $R^n$ in the target space  $\mu$. If $\mu (g(E_k)) = 0$ for all k then $\mu(g(E)) = 0$ as a countable union of measure zero sets is measure zero.  
Since $E_k$ has measure zero, for any $\delta > 0$ there is a cover of $E_k$ with open balls $\{B(x_i,r_i)\}_{i=1}^m$ such that their total measure is less than or equal to $\delta$. So,
$$ 
E_k\subset \bigcup\limits_{i=1}^m B(x_i,r_i) 
$$

If there is an $x\in E_k$ such that one of the m balls that covers $E_k$ covers x: $|x-x_i|\leq r_i$.  Thus if a function is Lipshitz $|y-g(x_i)|\leq \lambda_k r_i$ where $\lambda_k$ is the Lipshitz constant.  Thus
$$ 
\mu(g(E_k))\leq\mu\left(\bigcup\limits_{i=1}^m B(g(x_i),\lambda_k r_i)\right)\leq \lambda^n_k \delta. 
$$

As the radius of the balls go to zero as $\delta$ goes to zero and the entire set is still covered.  Thus $\mu(g(E_k)) = 0$, proving $\mu(g(E)) = 0$ $\blacksquare$ \newline

\textbf{Lemma 3}: \textit{If a function J maps a random variable (the base distribution) $X$ to another random variable $Y$ (the target), and both random variables are subset of $R^m$, a sequences of functions $J_n$ converges to $J$ pointwise, then the random variables $Y_n = J_n(X)$ converges in distribution to $Y = J(X)$.}  \newline 

\textbf{Proof}: (from \citep{huang2018neural}) Let $h$ be any bounded continuous function on $R^n$. $h$ composed with $J_n$ converges pointwise to $h \circ J$ as compositions of continuous functions converge to their compositions. Since $h$ is bounded, by the dominated convergence theorem $E[h(Y_n)] = E[h(J_n(X))] \to E[h(J(X))] = E[h(Y)]$. Since this result holds for any continuous function by the Portmanteau theorem this implies $Y_n \to Y$.  $\blacksquare$ \newline

Now that these lemmas have been proven, I will discuss an individual bijector. I will first show why it is bijective in event space (ie samples) and then move to probability space and show why it also is a valid change in measure and how to calculate the change of variables. Thus, if one knows the pdf of the base distribution, one also knows the distribution of the target distribution and vice versa.  Lastly, I will discuss how to set up the model in a particular way, so that it is able to universally approximate any smooth bijective mapping between two random variables, resulting in being able to map any continuous base distribution to any continuous target distribution, which allows this density estimator to have a ``non-parametric” universal approximator of distributions.  

I will work with the autoregressive bijectors discussed up to this point:

$$y^j_i = w(y^i_1,...y^i_{i-1})*\sigma(\alpha(y^i_1,..y^i_{i-1})*y^i_i + \beta(y^i_1,...y^i_{i-1}))$$

At this point, I will work in event space, so its not important to think about probability distributions but rather a mapping between two sets of numbers in $R^n$ (ie samples from a multivariate distribution). A bijector will be shown to be a bijection and so the dimensionality between samples in the base space and the target space should be the same\footnote{This can be relaxed but is unimportant for this paper}. Call this n. The autoregressive flow orders the input and the output as so $y^i = \{y^i_1, y^i_2...y^i_N\}, y^j = \{y^j_1, y^j_2...y^j_N\}$. The name autoregressive comes from the fact that each output variable is only nonlinearly conditioned on input variables with a smaller index and $y^j_1$ is just the identity of $y^i_1$. For example, the psuedo-parameters in the $y^j_5$ equation is only a function of $y^i_1, y^i_2, y^i_3, y^i_4$.  $y^i_5$ only enters in an affine manner so is easily invertable. See the section \ref{flows} for a more detail explanation of the architecture of a neural autoregressive normalizing flow. 

After laying down the intuition, it is time to prove a bijector is a bijection. \newline
\textbf{Theorem 1}: \textit{A bijector learns a bijection between the input and output space.}\newline

\textbf{Proof}: This will be done by proving three different elements are bijections and then the $k$ sum of these elements are bijections.  First $y^j = \alpha(y^i_1,...y^i_{n-1})*y^i_n + \beta(y^i_1,...y^i_{n-1})$ is a bijection. Then $\sigma(.)$ is a bijection and finally $y^j = w(y^i_1,...y^i_{n-1})*\sigma(.)$ is one. In order to prove that the first relationship is onto, first recognize the identity function is onto as $y^j_1 = y^i_1$. Thus one can recover $y^i_1$ knowing $y^j_1$. Now proceed via induction. In order to get $y^i_n$ from $y^j_n$ assume one knows $y^i_{1:n-1}$. Thus given any $y^i_n$, $\alpha$ and $\beta$ are fixed as a function of earlier $y^i_{1:n-1}$’s and thus are constant values with respect to $y^i_n$. Then given a $y^j_n$, one can recover a corresponding $y^i_n$ that will produce it via the equation $(y^j_n-\beta)/\alpha$. To prove that this function is 1-1 proceeds in the same way.  The identity is 1-1 and if $y^j_n = {y'}^j_n$, then $y^i_n = {y'}^i_n$, again by the fact that affine functions are 1-1, $\alpha$ and $\beta$ are constants if one knows all previous elements of $y^i$. Then one extends this to all dimensions with induction in the same way as onto was proved. Proving $\sigma(y')$ is a bijection is by definition. $\sigma$ is chosen to be a bijective nonlinearity.  Proving the final step, $y^j = w(y^i_1,...y^i_{n-1})*\sigma(.)$, as this again is linear function in $\sigma(.)$ which is a subset of affine functions and in the first set of equations, I proved that affine functions are bijections, thus linear functions are bijections too. The final step involves proving the bijectiveness of $y^j_i$ is made up of a sum of $k$ of these equations. However, it is not true that a sum of bijections is a bijections.  For example take $f = x$ and $g = -x$.  One condition for a sum of bijections to be a bijection is for the bijections to be monotonic. In lemma 1, I prove that if the weights on the neural network in a neural autoregressive bijector are all positive and that the input (e.g. $y^i$ as opposed to $y^j$) distribution is positive, this will make the model monotonic. Proceeding, given two monotonic continuous bijectors $f$ and $g$, the sum is onto. Define $h$ as $f+g$.  $h$ is continuous as a sum of continuous functions is continuous.  Given a value $y$ in the interior of the range of $h$ there exists a value of $h|_v > y$ and a value of $h|_w < y$.  Since $h$ is continuous, by the intermediate value theorem it takes on a value of $y$.  Likewise to prove 1-1, $h$ is also monotonic as a sum of monotonic functions is monotonic. Thus, it can only take on that value of $y$ exactly once. One can extend the proof of two monotonic bijections to arbitrary monotonic bijections by using induction. Combining the neural autoregressive bijector with a bijector that permutes the indices of $y^i_n$ allows for arbitrary conditioning relationships. Now with a stack of bijectors, where the output of one bijector is the input of another, this is also a bijection as a composition of bijections is a bijections. Thus a normalizing flow is a bijection.  
$\blacksquare$ \newline

Proving bijectiveness, now I will work in the underlying density space/the space of measures and move away from just viewing the flow as a mapping of realizations of random variables. I will now proceed to show that a normalizing flow is C-Lipschitz when the total domain is restricted to a set which has probability mass $1 - \delta$ for arbitrarily small delta. This will allow me to show that the input of a normalizing flow absolutely continuous with respect to the output and vice versa. This is Lipsitz bounded by a constant $(\max_{|\alpha|,|\beta|} |\alpha| + |\beta|)^N$. Define the $max_{|\alpha|,|\beta|}$  of $\alpha$ and $\beta$ as the maximum value these psuedo-parameters take on the compact set $V(1-\delta)$. The boundedness of the normalizing flow is just $(\max_{|\alpha|,|\beta|} |\alpha| + |\beta|)^N$ because each normalizing flow defines an affine relationship $\alpha(y^i_1,…y^i_{n-1})*y^i_n + \beta(y^i_1,…y^i_{n-1})$, which is bounded itself. Then, $\max_{\alpha,\beta} \alpha + \beta$ is the largest value of any $\alpha$ and $\beta$ all the bijectors of the compact set take. Raising it the the power $N$ corresponds to the number of times the bijector is composed with itself. Thus, the Lipschitz constant cannot be greater than this number. Likewise the reverse direction finds the max for the inverse values for alpha. This implies that the function is C-Lipschitz over any particular compact set that includes our data up to probability $\delta$.  \newline

\textbf{Theorem 3}: \textit{The chain of neural autoregressive bijectors along with a bijector that permutes the indices of inputs, learns a valid change in measure of an input base distribution to an output target distribution and the pdf of the target distribution can be derived using the standard change of variable formula: $q(y^z) = q(y^a)|det (df/dy^a)|^{-1}$}\newline

\textbf{Proof}: I will now use Lemma 2 which proved that any Lipschitz functions preserves sets with measure zero. This implies that measures of both spaces are absolutely continuous with respect to one another. Additionally Lemma 3 which proves that if a sequence of flows $f_n$ converges to $f$ point-wise, the mapping of random variables $f_n(X)$ converges to $f(X)=Y$. Thus in both directions there exist a Radon Nikodym derivative and the Radon Nikodym derivative of the transformation of one random variable transformed by a bijector follows the convenient change of variables formula:
$q(y^z) = p(y^a = f^{-1}(y^z))|det (df^{-1}/dy^a)|$ evaluated at a point $y^z$.  
where $f$ maps $y^a$ to $y^z$ ($f(y^a) = y^z$).  This can be even more conveniently expressed as:
 $q(y^a)|det (df^{-1}/dy^a)| = p(y^a)|det (df/dy^a)|^{-1}$ 
This relationship can be shown by using the inverse function theorem and the recognition that the determinant of an inverse is the inverse of the determinant: $|det (df^{-1}/dy^a)| = |det ((df/dy^a)^{-1})|=|det ((df/dy^a))^{-1}|=|det (df/dy^a)|^{-1}$, with derivatives all evaluated at a point $y^a$. 

Now I have shown given a sample from $y^z$ and a given distribution like $y^a$, how to recover the sample points in $p(y^a)$ that correspond to the points $q(y^z)$. First one inverts the bijector to get $y^a$. Then one can evaluate the base distribution on $y^a$. Finally, to calculate the likelihood of $q(y^z)$, one uses the change of variables formula to get the pdf of $y^z$ under the bijector. One then can optimize the bijector using maximum likelihood and $q(y^z)$ as the likelihood of $y^z$.  

A normalizing flow is just a stack of bijectors on top of one another so the output of one bijector becomes the input of the next bijector.  Composition of bijections are also bijections and is invertible.  Likewise a composition of Lipschitz functions is still Lipschitz and so this flow also is a valid change of measure. This proves that a normalizing flow, which is a stack of bijectors, is a valid change of measure whose density can be evaluated using the Radom Nikodym formula: $q(y^z) = p(y^a)|det (df/dy^a)|^{-1}$.  $\blacksquare$ \newline

Finally I will show that a certain construction of normalizing flows are universal approximators of multivariate real probability distributions. This means the give a suitably chosen set of parameters the normalizing flow will converge in distribution to any continuous probability measure. In the proof, mirroring universal approximation proofs in general, I will focus on summing up an infinite number of bijectors, proving universal approximator properties for extremely wide normalizing flows, even though in practice deep normalizing flows are more often used.  

Since the variables are conditioned only on previously indexed variables, of course a sum of bijectors all with the same conditioning structure between variables cannot approximate any dependency relationships between a target variable and a base variable. In practice this can be avoided by summing or stacking multiple bijectors and permuting the indices of the input, changing which variables are conditioned on which other variables. 

\subsection{Normalizing Flows are Universal Approximators}

The proof that normalizing flows are universal approximators will proceed in a series of steps following \citet{huang2018neural}. Initially, I will show that the flow can approximate any monotonic function. The first sub-step is to show show that a sum of step functions are universal approximators of monotonic functions. Then I will show that if we make the nonlinearity $\sigma$ a logit transformation, one can modify the logit transform to be arbitrarily close to a given step function. Then the proof will show that a neural network that approximates the optimal parameters $w$, $\alpha$, $\beta$ of the logit function can universally approximate any monotonic function. Finally, any continuous probability distribution can be mapped to any other probability distribution with a monotonic function. This will be made formal by showing monotonic functions can map any pdf to and from the multivariate uniform distribution.   

In lemma 4-6 the domain and range of the two functions are between two arbitrary constants $(r_0,r_1)$\newline

\textbf{Lemma 4}: \textit{Step functions universally approximate monotonic functions}\newline

Define: 
$$
    S^*(y) = \sum_k w_k*s(y-b_j)
$$

Where $s$ is the step function that steps at $y = b_j$

Then the lemma states that for any continuous and monotonically increasing function $S : [a,b] \to [0,1]$ and given any $\epsilon > 0$ there exists an $S^*(n)$ for some n such that $|S^*(y)-S(y)| < \epsilon \forall y \in [a,b]$.  Abbreviating $s_j(y) = s(y-b_j)$, choose an $n = \left \lceil{1/\epsilon}\right \rceil$ with the ceiling function and divide the range $[0,1]$ into n+1 evenly spaced intervals: $[0,z_1),[z_1,z_2)...[z_n,1]$.  For each $z_j$, there is a corresponding $y_j$ value that corresponds to $y_j = S^{*-1}(z_j)$, since $S^*$ is strictly monotonic.  Now find the step function so all the $S^*(y_j) = z_j$, which will mean the error is at most $\epsilon$ as the difference in $z_j$ is at most $\epsilon$.  To do this ensure that $s_j(y)$ steps at $y_j$ by setting $b_j = y_j$. Then set the weights $w_j$ so that $w_j$ makes up the difference between $z_{j-1}$ and $z_j$.  This is equivalent to solving $Sw = t$ where $S$ is the lower triangular matrix of ones that represents which steps have activated and $t$ is the vector of $[0, z_1, z_2...z_n,1]$. The distance between $t_j - t_{j-1} = 1/(n+1)$, thus: 
\begin{align}
    |S*(y) - S(y)|  &= |\sum_j s_j(y)(t_j-t_{j-1})-S(y)| \\
                    &= |1/(n+1)*\sum_j s_j(y) - S(y)| \\
                    &\leq 1/(n+1) < \epsilon
\end{align}
 $\blacksquare$\newline
 
\textbf{Lemma 5}: \textit{I will move on to proving the same thing with sigmoid functions of the form}:
 $$
    S'(y) = \sum_k w_k*\sigma((y-b_k)/\tau_k)
 $$

\textit{also can universally approximate arbitrary monotonic functions.}\newline 

Note this can easily be translated into a form similar to our bijector by setting $\alpha_k = 1/\tau_k$, $beta_k = b_k/\tau_k$, and $w_k = w_k$.  Note at this point $\alpha$, $\beta$ and $w$ are constants and not functions. I will deal with making them functions in the next lemma. For any monotonic function, approximate the function using a step function up to a factor of $\epsilon/3$, using the previous lemma. Define $S'(y)$ such that $w_k = w_j$ and $b_k = b_j$ where the $j$ index refers to the step function parameters ($S*$) and the $k$ index refers to the parameters in the sum of sigmoids ($S'$). Then choose $\tau = \tau_k = \dfrac{\kappa}{\sigma^{-1}(1-1/(n-1))}$ where $\kappa = \min_{l\neq l'}b_l-b_{l'}$. Then $|\sigma((y-b_k)/\tau) - s(y-b_k)|$ is at most $.5$ if $b_{k-1}<y<b_{k+1}$ as the logit is around $.5$ right at $b_k$ where the step steps from 0 to 1 and since the size of the steps are of magnitude $\epsilon/3$, the error is at most $.5*\epsilon/3$. However, if $y$ is not in this range then $sigma(y-b_k/\tau) = \sigma(y-b_k/(\min_{l\neq l'}b_l-b_{l'})*\sigma^{-1}(1-1/(n-1))$ which implies that the distance from the step is at most $1/(n-1)$ for a 0 to 1 step function.  Putting this together you have $\epsilon/3$ from the sum of the two one halves, then $\epsilon/3  = (n-1)/(n-1)*\epsilon/3$ from each other sigmoid that is not within $b_{k-1}<y<b_{k+1}$, which gives $2/3\epsilon$ from the logit approximation to the step function and $\epsilon/3$ from the step function approximation to the monotonic function. Using the triangle inequality: $ |S'(y)-S(y)| < |S'(y)-S*(y)|+|S*(y)-S(y)| < 2/3\epsilon + \epsilon/3 = \epsilon$ $\blacksquare$\newline

\textbf{Lemma 6}: \textit{Let the base and target space be multivariate and $\alpha$, $\beta$, and $w$ now be the outputs of neural networks conditioned on previous data.  Bijectors parameterized like this}:
\begin{equation}
    y^j_n = \sum_k w_k(y^i_1,...y_{n-1})*\sigma(\alpha_k(y^i_1,...y_{n-1})*y^i_n + \beta_k(y^i_1,...y^i_{n-1}))    
\end{equation}

\textit{universally approximates monotonic functions.}\newline

I will prove this theorem by writing the bijectors in this form:
\begin{equation}
    y^j_n = \sum_k w_k(y^i_1,...y^i_{n-1})*\sigma(\dfrac{y^i_i - b_k(y^i_1,...y^i_{n-1})}{\tau_k(y^i_1,...y^i_{n-1})})    
\end{equation}

Since $n$ indexes all the outputs of the bijector, I will proceed by proving the theorem for univariate case for each $y^j_n(y^i_1...y^i_n)$ and then extending the theorem for the multivariate function (eg the distance all $n$ have from a multivariate monotonic function is smaller than a given $\epsilon$) is relatively easy. 

Proving this theorem requires showing there is a sequence of functions $y^j_n(y^i_1..y^i_n)$ such there exists a $k<K$ summed bijectors in $w_k(y_1,...y_{i-1})$, $b_k(y_1,...y_{i-1})$, and $\tau_k(y^i_1,...y^i_{n-1})$, that the error $|y^j_n(y^i_1...y^i_n)-S(y_n)|$ is smaller than any $\epsilon > 0$. Here $y^j$ is the bijector and $S$ is the monotonic function that needs to be approximated.  

Since the fixed parameters can approximate any monotonic function arbitrarily well, I have from lemma 5:
$$
    |S'(y^i_n)-S(y^i_n)| < \epsilon/2
$$

I also know from \citet{cybenko1989approximation}, that a wide enough neural network can approximate any function. Choosing a function that matches the parameters $S'(y_i)$ with low enough error, I know there exists a neural network for the parameters $\alpha_k(.)$, $w_k(.)$, and $\tau_k(.)$ in $y^i_n$ such that:
$$
    |y^j_n(y^i_1...y^i_n)-S'(y^i_n)| < \epsilon/2
$$

Now this distance bound is only true because the bijector function $y^j$ is a bounded and continuous and so you can use the dominated convergence theorem to turn the the convergence of the sequence of parameters (e.g. $\lim_{l\to\infty}\alpha_k^l(.) \to \alpha_k$ etc), to a statement about the convergence of a sequence of functions: $\lim_{l\to\infty}\{y^j_n\}^l \to S'(x_n)$

Then the proof in the univariate case is finished by using the triangle inequality:
\begin{align}
    |y^j_n(y_1...y_n) - S(y^i_n)| &< |y^j_n(y_1...y_n)-S'(y^i_n)| + |S'(y_n)-S(y_n)| \\
    &=\epsilon/2+\epsilon/2 = \epsilon
\end{align}

The proof in the multivariate case is completed by recognizing that for each univariate case there is a $K_i$ number of summed bijectors such that when $k_i>K_i$, $|y^j_n(y_1...y_n) - S(y^i_n)| < \epsilon$.  Then just choose $k = \max_{k_i} $ and then $||y^j(.)-S(y_i)||_\infty < \epsilon$ $\blacksquare$\newline

Now that I've demonstrated that a single very wide collection of bijectors is enough to approximate any monotonic function, it's clear stacking many wide bijectors can only improve model flexibility for a given width $k$. What is left to show is that monotone functions are enough to allow a normalizing flow to map any continuous base distribution to any continuous target distribution. The proofs will show that any distribution can be mapped to a multivariate uniform distribution with a normalizing flow. Likewise any distribution can be mapped from a uniform distribution. Since both are true, then it is clear that a flow can map any distribution to any distribution. These proofs are similar in spirit to Sklar's theorem for copulas, as copulas share similar proprieties as normalizing flows. The proof is complementary to econometric proofs for identification in non-separable models in the sense that it can in some non-rigorous way to be considered an existence proof to \citet{matzkin2003nonparametric} partial uniqueness identification argument.  \newline

\textbf{Lemma 7}:Normalizing flows transform multivariate uniform random variables to arbitrary random variables. \textit{The neural network parametrized by} $$
G(y^i)_t = \sigma^{-1}(\sum_k w_k(y^i_1,...y^i_{n-1})*\sigma(\alpha_k(y^i_1,...y^i_{n-1})*y^i_n + \beta_k(y_1,...y_{n-1})))
$$
\textit{can transform a base distribution ($Y \sim Unif((0,1)^m)$ of an m dimensional uniform random variables to any desired m dimensional random variable. Call this target distribution $Z \in R^m$. This convergence is in distribution to the desired random variable}\newline

This equation for $G_n$ is the same equation as in lemma 6, with $\sigma$ being a sigmoid/logit distribution. The one exception is the addition of the inverse sigmoid. \newline

\textbf{Proof}: Let $Z_t$ be a random vector in $R^m$ and assume $Y$ has strictly positive and continuous density. Take $Y$ as the m dimensional uniform. Then there is a sequence of functions $\sigma^{-1} \circ \{y^z\}_t(y^a)$ where $\{y^z\}_t$ is the normalizing flow as discussed in lemma 6. Given an arbitrary ordering of the multiple CDFs of $Z^t$, defined as $F_t(Z_m<Z_m|y_1...y_{m-1})$, according to theorem 1 in \citet{hyvarinen1999nonlinear}, $F(Z)$ is distributed uniformly in the m dimensions and $Z$ is independent of the $y$'s. Then the inverse CDF function $G_t$ is a monotonic function, as is $\sigma \circ G_t$.  $G_t(F_t(Z_m<z_m|y_{1:m-1})|y_{1:m-1}) = Z_m$. Multivariate CDFs have an upper triangle dependency relationship. $G_t$ is mapping from $F_t$ which is a uniform distribution to the realizations of the random variables $y_t$. This by definition is a monotonic mapping. By lemma 6, $\sigma \circ G_t$, can be approximated arbitrarily well by a sequence of bijectors, $\{y^z\}_t(y^a)$. The $\sigma$ is necessary because the preimage of $\{y^z\}_t(y^a)$ in lemma 6 is $(r_0,r_1)$, which is only a bounded subspace of $R^n$.  Thus, $G_t = \sigma^{-1} \circ \{y^z\}_t(y^a)$ converges uniformly, thus pointwise, to $G$, as $\sigma^{-1}$ is continuous.  Since $G_t$ converges to $G$ and $G(Y) = Z$, by lemma 2 we have $G_n \to Z$ in distribution.  $\blacksquare$ \newline

\textbf{Lemma 8}: Given a correct parameterization, a normalizing flow that maps from arbitrary distributions can converge in distribution to an multivariate uniform (0,1) distribution.\textit{Let $Y$ be a positive and continuous random vector in $R^m$. Let $Z$ be distributed according to the multivariate uniform distribution.  The there exists a sequence of functions $\{y^z\}_t(y^a)$ parameterized as in lemma 6, that converges in distribution to $Z$.}\newline  

\textbf{Proof}: Define the CDF for all y dimensions as: $z_1 = F_1(y_1|0)$ and $z_m = F_t(y_m|z_1...z_{m-1})$. To be clear $y_t = Pr(Y_m<y_m|z_1...z_{m-1})$, which again is a multivariate CDF with an upper triangular form. Again due to the theorem in \citet{hyvarinen1999nonlinear}, the distribution of the CDF is an n dimensional uniformly distribution.  Then since lemma 6 shows that there exists a monotonic function, $\{y^z\}_t(y^a)$ that converges uniformly, thus pointwise, to $F$ and $F(Y) = Z$, we have $\{y^z\}_t \to Z$ in distribution.  $\blacksquare$\newline

\textbf{Theorem 9}: \textit{Bijectors in the form,}
$$
H(y)_t = \sigma^{-1}(\sum_k w_k(y^i_1,...y^i_{n-1})*\sigma(\alpha_k(y^i_1,...y^i_{n-1})*y^i_n + \beta_k(y^i_1,...y^i_{n-1})))
$$

\textit{can map arbitrary distributions to arbitrary distributions arbitrarily closely in distribution} \newline  

\textbf{Proof}: The intuition here is to map any distribution to a uniform distribution by lemma 7 and then the uniform to any other distribution by lemma 8. Again from theorem 1 from \citet{hyvarinen1999nonlinear} in lemma 8 $F(Y)$ is uniformly distributed so is $G(F(Y)) = Z$, from lemma 7.  Since both functions are monotonic, a composition of the two monotonically increasing functions is also a monotonic increasing function and can be mapped via a bijector.  According to lemma 7, there exists a sequence of functions, $\{y^z\}_t(y^a)$ that converge to $\sigma \circ G \circ F$ uniformly, thus pointwise. Then there exists a $K_n = \sigma^{-1} \circ \{y^z\}_t(y^a)$ that converges pointwise to $G \circ F$, by monotonicity of $sigma^{-1}$.  Then by lemma 2 we have $K_n \to Z$ in distribution.  $\blacksquare$
\newpage

\noindent \textsc{Department of Economics, University of Michigan, Ann Arbor}\newpage 

\newpage 

\end{document}